\documentclass{article}
\usepackage[T1]{fontenc}
\usepackage{mathtools}
\usepackage[normalem]{ulem}
\usepackage{amsmath,amsthm}
\usepackage{amssymb}
\usepackage{slashed}
\usepackage{multirow}
\usepackage{fullpage}
\usepackage[bottom]{footmisc}
\usepackage{longtable}
\usepackage{mathrsfs}
\usepackage{empheq}
\usepackage{comment}
\usepackage{physics}
\usepackage{dsfont}
\usepackage{xcolor}
\usepackage{tocloft}
\usepackage{titlesec}
\usepackage{stmaryrd}
\usepackage{bm}
\usepackage{bbm}
\usepackage{float}

\usepackage{jheppub}

\makeatletter
\DeclareRobustCommand{\loplus}{\mathbin{\mathpalette\dog@lsemi{+}}}

\newcommand\sbullet[1][.4]{\mathbin{\vcenter{\hbox{\scalebox{#1}{$\bullet$}}}}}

\newcommand{\defeq}{\vcentcolon=}

\newcommand{\dog@lsemi}[2]{\dog@semi{#1}{#2}{270,90}}
\newcommand{\dog@semi}[3]{%
  \begingroup
  \sbox\z@{$\m@th#1#2$}%
  \setlength{\unitlength}{\dimexpr\ht\z@+\dp\z@\relax}%
  \makebox[\wd\z@]{\raisebox{-\dp\z@}{%
    \begin{picture}(1,1)
    \linethickness{\variable@rule{#1}}
    \roundcap
    \put(0.5,0.5){\makebox(0,0){\raisebox{\dp\z@}{$\m@th#1#2$}}}
    \put(0.5,0.5){\arc[#3]{0.5}}
    \end{picture}%
  }}%
  \endgroup
}
\newcommand{\variable@rule}[1]{%
  \fontdimen8  
  \ifx#1\displaystyle\textfont3\else
    \ifx#1\textstyle\textfont3\else
      \ifx#1\scriptstyle\scriptfont3\else
        \scriptscriptfont3\relax
  \fi\fi\fi
}
\makeatother
\usepackage{tikz-cd}

\usetikzlibrary{arrows.meta,calc}

\usepackage{cleveref}
\usepackage{pict2e}
\usepackage{diagbox}
\usepackage{calc} 

\newcommand\Lie{\pounds}
\newcommand{\CC}{\mathbb{C}}
\newcommand{\RR}{\mathbb{R}}
\newcommand{\HH}{\mathbb{H}}
\newcommand{\ZZ}{\mathbb{Z}}

\newcommand{\SSS}{\mathbb{S}}
\newcommand{\dS}{\mathfrak{diff}(S)}

\newcommand{\g}{\mathfrak{g}}
\newcommand{\h}{\mathfrak{h}}

\newcommand{\gl}[1]{\mathrm{GL}(#1,\mathbb{R})}
\newcommand{\spl}[1]{\mathrm{SL}(#1,\mathbb{R})}
\newcommand{\Liegl}[1]{\mathfrak{gl}(#1,\mathbb{R})}
\newcommand{\Liespl}[1]{\mathfrak{sl}(#1,\mathbb{R})}

\newcommand{\so}[1]{\mathrm{SO}(#1)}

\newcommand{\su}[1]{\mathrm{SU}(#1)}

\newcommand{\mh}{\mathfrak{h}}

\newcommand{\chkM}{{\color{red} \,\checkmark\kern-5pt{}_{M}}}

\newcommand{\ee}{\end{equation}}
\newcommand{\bea}{\begin{eqnarray}}
\newcommand{\eea}{\end{eqnarray}}

\newcommand{\dM}{\mathfrak{diff}(M)}

\newcommand{\End}{\text{End}}
\newcommand{\Aut}{\text{Aut}}

\newcommand{\pair}[2]{\langle#1,#2\rangle}
\newcommand{\Ad}[1]{\mathrm{Ad}_{#1}}
\newcommand{\ad}[1]{\mathrm{ad}_{#1}}
\newcommand{\cAd}[1]{\mathrm{Ad}^*_{#1}}
\newcommand{\cad}[1]{\mathrm{ad}^*_{#1}}
\newcommand{\el}[1]{\bm{#1}}
\newcommand{\elg}[1]{\bm{#1}}
\newcommand{\mommap}{\mathcal J}
\newcommand{\QCSop}{\hat{\mommap}^{\mathrm{QCS}}}
\newcommand{\DL}{\tilde L}
\newcommand{\lbr}{\llbracket}
\newcommand{\rbr}{\rrbracket}
\newcommand{\lga}{\triangleright}

\newcommand{\Lieb}[1]{[#1]_{\mathrm {Lie}}}
\newcommand{\bbd}{\dd}

\newcommand{\Fock}{\mathfrak F}

\newcommand{\SISSA}{\affiliation{$^1$~SISSA, International School for Advanced Studies, \\
via Bonomea 265, 34136 Trieste, Italy}}
\newcommand{\InfnTS}{\affiliation{INFN, Sezione di Trieste,\\
via Valerio 2, 34127 Trieste, Italy}}
\newcommand{\IFPU}{\affiliation{IFPU, Institute for Fundamental Physics of the Universe,\\
via Beirut 2, 34014 Trieste, Italy}}
\newcommand{\ncbj}{\affiliation{$^2$~National Centre for Nuclear Research, Pasteura 7, 02-093 Warsaw, Poland}}

\title{Orbit method for Quantum Corner Symmetries}

\author{Giulio Neri$^1$, Ludovic Varrin$^2$}

\date{\today}

\SISSA
\InfnTS
\IFPU
\ncbj

\emailAdd{gneri@sissa.it}
\emailAdd{ludovic.varrin@ncbj.gov.pl}

\abstract{
    The classification of the unitary irreducible representations of symmetry groups is a cornerstone of modern quantum physics, as it provides the fundamental building blocks for constructing the Hilbert spaces of theories admitting these symmetries. In the context of gravitational theories, several arguments point towards the existence of a universal symmetry group associated with corners, whose structure is the same for every diffeomorphism-invariant theory in any dimension.
    Recently, the representations of the maximal central extension of this group in the two-dimensional case have been classified using purely algebraic techniques. In this work, we present a complementary and independent derivation based on Kirillov’s orbit method.
    We study the coadjoint orbits of the group $\widetilde{\spl{2}}\ltimes\HH_3$, where $\HH_3$ is the Heisenberg group of a quantum particle in one dimension. Our main result is that, despite the non-abelian nature of the normal subgroup in the semidirect product, these orbits admit a simple description. In a coordinate system associated with modified Lie algebra generators, the orbits factorize into a product of coadjoint orbits of $\spl{2}$ and $\HH_3$. The subsequent geometric quantization of these factorized orbits successfully reproduces the known representations.    
}
\arxivnumber{}

\begin{document}

\maketitle

\section{Introduction}
\label{sec.Intro}

Symmetries have played a fundamental role in physics since the pioneering contributions of Lagrange, Hamilton, and Jacobi \cite{lagrange1811, lagrange1815, hamilton1834,hamilton1835,jacobi1866}. The modern treatment of symmetries in theoretical physics originates from the seminal work of Noether in 1918 \cite{noether1918}. Although initially motivated by the challenge of energy conservation in the then recent theory of general relativity, Noether's results have had a far broader impact. Today, her theorems stand as a central tool in quantum field theory.

Noether's first theorem states that for every continuous symmetry of a system, there exists an associated conserved quantity, known as the Noether charge. One can also take the opposite perspective and ask: given a set of symmetries, is there a system that features them? In the quantum case, the answer come from representation theory. Specifically, given a system with a symmetry group, any quantum state must lie in an irreducible unitary representation of that group. By combining such representations, one can construct quantum systems with the desired symmetry properties. This idea is at the core of Wigner's famous classification of the irreducible unitary representations of the Poincaré group, which underlies the modern formulation of relativistic quantum field theory~\cite{Wigner:1939cj}.
In the classical case, the analogous  question remained unsolved for some time: can one construct a system whose phase space encodes a given symmetry group? The answer came with the work of Kirillov, Kostant, and Souriau on coadjoint orbits \cite{Souriau1958,kirillov_unitary_1962,souriau1970structure,Kostant1970,kirillov_geometric_1990,KirillovLectures}. The statement is that the classical phase space of a system is naturally expressed in terms of coadjoint orbits of its symmetry group. Moreover, Kirillov's orbit method postulates a correspondence between coadjoint orbits and unitary irreducible representations, thereby bridging the classical and quantum pictures.

How can these ideas be applied to gravity? Is there an analogue of Wigner’s classification for the symmetries underlying gravitational theories? An important difference from the Poincaré group is that the symmetries of gravity are local, so one must turn to Noether’s second theorem to address these questions. This theorem states that any local symmetry corresponds to a constraint, which in turn reflects a redundancy in the equations of motions. To be more precise, the conserved charges of Noether's first theorem are associated with codimension 1 objects -- the Noether currents -- that yield the charge once integrated over a codimension 1 Cauchy hypersurface. When the symmetry is local, the second theorem states that such a current is an exact $(d-1)$-form on-shell of the equations of motion and, by Stokes theorem, its integral localizes on the boundary of the Cauchy hypersurface (a codimension 2 surface, called a \textit{corner}). In the absence of boundaries, such charge identically vanishes. 
As this charge does not correspond to an observable, the associated symmetry is not a physical one but is instead a gauge redundancy. States connected by such a transformations have to be identified. However, when boundaries are present, some of the charges become non-vanishing and one says that the corresponding symmetry is physical.

In the gravitational context, corner charges first appeared in the work of Regge and Teitelboim~\cite{regge_role_1974}. More recently, the influential work of Donnelly and Freidel~\cite{donnelly_local_2016} sparked renewed interest in gravitational corner charges, leading to an extensive study of the topic~\cite{speranza_local_2018, freidel_edge_2020, donnelly_gravitational_2021, Speranza:2022lxr, ciambelli_isolated_2021, ciambelli_embeddings_2022,freidel_extended_2021, geiller_extended_2020, Carrozza:2022xut,Freidel:2023bnj,ciambelli_universal_2023,Langenscheidt:2024nyw, ciambelli_quantum_2024}. In particular, it was found that the corner charges of any gravitational theory realize a (possibly improper) subset of the \textit{universal corner symmetry} (UCS) algebra \cite{ciambelli_universal_2023} 
\begin{equation}\label{eq:ucsalgebra}
    \mathfrak{ucs} = \mathfrak{diff}\qty(S) \loplus \qty(\mathfrak{gl}\qty(2,\RR)\loplus \RR^2)^S,
\end{equation}
where $S$ denotes the corner under consideration; its presence in the exponent indicates that there exists a copy of the algebra inside the parentheses for each point of the corner. The algebra $\dS$ has a clear geometric interpretation, while $\Liegl{2}$ corresponds to boosts and dilatations in the normal plane to the corner, and the $\RR^2$ factor corresponds to translations in the two normal directions. The presence of the semidirect sum will be clarified in Section~\ref{sec.UCS}.
As we will see there, the construction of the UCS algebra only relies on properties of vector fields, making it ubiquitous in theories with diffeomorphism invariance. Because of its universality, the algebra~\eqref{eq:ucsalgebra} lies at the core of the \textit{corner proposal}. In this approach, the group associated with the UCS algebra should be seen as the gravitational analogue of the Poincaré group in relativistic quantum field theory. More specifically, it is conjectured that quantum gravitational states are organized according to the irreducible unitary representations of the UCS group or one of its subgroups.

The next step of the program is clear: find the irreducible unitary representation of the UCS. Unfortunately, the algebra \eqref{eq:ucsalgebra} is infinite-dimensional and complicated by the semidirect sum structure with diffeomorphisms, so that a full classification of its representations is not known yet. One possibility is to first tackle a simplified version of the problem by focusing on the group
\begin{equation}
    \mathrm{UCS}_2 = \mathrm{GL}\qty(2,\RR) \ltimes \RR^2,
\end{equation}
which can be seen either as a restriction to the two-dimensional case (when the diffeomorphism algebra trivializes) or as a per-point analysis of the corner in the full-dimensional case.
Furthermore, the determinant part $W$ of the general linear group -- $\gl{2}=W\ltimes \spl{2}$ -- is not realized at finite-distance corners in all theories that have been explicitly studied so far. It is believed to be associated instead with symmetries at asymptotic corners, where it appears as the Weyl factor of the generalized Weyl-BMS group \cite{freidel_weyl_2021}. Therefore, we can restrict to the so-called \textit{extended corner symmetry} (ECS) group in two dimensions
\begin{equation}
    \mathrm{ECS}_2 = \spl{2}\ltimes \RR^2.
\end{equation}
The study of representations of this group and their applications to the description of local subsystems was initiated in~\cite{ciambelli_quantum_2024} and continued in~\cite{varrin_physical_2024}. A key observation in the study of representation theory is that, since states in a Hilbert space are not directly observable \textit{per se}, one must to allow for projective representations. In turn, this calls for a classification of all possible central extensions of the symmetry group. In these works, it was found that the group admits a unique central extension in the translational sector (in addition to the usual lifting to the universal cover $\spl{2}\to\widetilde{\spl{2}}$). The resulting maximally extended group has been dubbed \textit{quantum corner symmetries} (QCS) group:
\begin{equation}
    \mathrm{QCS}_2 = \widetilde{\spl{2}}\ltimes \HH_3,
\end{equation}
where $\HH_3$ denotes the three-dimensional Heisenberg group. The complete classification of unitary irreducible representations of the QCS has been recently obtained by one of the authors~\cite{varrin_physical_2024}. Given the relevance that these representations play in our understanding of quantum gravity (and particularly in the corner proposal), we want to provide a complementary and independent derivation.

In this paper, we therefore take up the work that was initiated in \cite{ciambelli_universal_2023} and set out to recover the unitary irreducible representations of the QCS via the orbit method. This derivation not only serves to corroborate the classification of~\cite{varrin_physical_2024}, but it also introduces a complementary, geometric perspective that we believe will be instrumental for future developments. In particular, it provides valuable insights about the particle dynamics interpretation -- analogously to how relativistic particles arise from the coadjoint orbits of the Poincaré group -- which is of great interest for the physics of the corner proposal.

Another advantage of following this approach is that, in higher dimension, the algebraic techniques used in~\cite{varrin_physical_2024} may no longer be sufficient, while the study of coadjoint orbits does not have any apparent obstruction. This generalization is essential to get a complete and correct picture. For instance, the appearance of a central element in the translation sector implies that translations no longer commute. While far from clear, this seems to suggest that spacetime itself becomes fuzzy at the quantum level, making it impossible to localize events with arbitrary precision. On the other hand, the authors cannot exclude at this point the possibility that the central extension is an artifact of the two-dimensional setting. Therefore, restoring the suppressed dimensions and reintroducing diffeomorphisms will be a crucial step. 

In the next section, we introduce the corner symmetry algebra of gravity and review the construction of its defining representation in terms of spacetime vector fields. We then discuss the extension of this algebra to the quantum case and summarize the results obtained so far in its representation theory.
In Section~\ref{Sec.Method}, we briefly present Kirillov's orbit method along with the basics of geometric quantization, which is the quantization technique we will use to produce Hilbert spaces from the orbits. In Section~\ref{Sec.Orbits}, we first review and generalize the application of the orbit method to the classical group of corner symmetries that was performed in~\cite{ciambelli_universal_2023}. Then, we apply the same method to the quantum case. We show that a careful analysis of its coadjoint orbits and their subsequent quantization reproduces -- modulo known subtleties and complications -- the representations derived in~\cite{varrin_physical_2024}. The explicit correspondence is established in Section~\ref{Sec.Reps}.

We adopt the following notational convention. Bold uppercase letters will denote vectors in the Lie algebra $\g$ of a certain Lie group $G$, while bold lowercase letters will denote (co-)vectors in the dual algebra $\g^*$. Therefore, once a pair of bases dual to each other has been chosen in these spaces, we write, for example, $\el{X}\in\g$ as $X^i\el{m_i}$ and $\el{m}\in\g^*$ as $m_i\el{X^i}$. This is consistent with the interpretation of $\el{X}^i$ as the linear functional $\el{X}\mapsto X^i$ and $\el{m}_i$ as the linear functional $\el{m}\mapsto m_i$, which is valid for finite-dimensional vector spaces. Furthermore, when many elements are involved and there is a proliferation of symbols, we will generally use greek letters for Lie algebra components and dual algebra basis elements, and latin letters for Lie algebra basis elements and dual algebra components.

\section{Corner symmetries in gravity}
\label{sec.UCS}

In this section, we review the concept of corner symmetries in gravity and some of the recent advances in their use and realization.

Given a differential manifold $M$ and the embedding $\phi:S\to M$ of a codimension 2 surface $S$ in that manifold, one can consider the maximal subalgebra of $\dM$ associated with $\phi(S)$. Explicitly, given local coordinates $(u^a,x^i)$ on $M$ close to $\phi(S)$, with $a=1,2$ and $i=1,\dots, d-2$, one can realize the embedding as
\begin{equation}
    \phi(\sigma)=(u^a(\sigma),x^i(\sigma))
\end{equation}
where $\sigma=(\sigma^1,\dots,\sigma^{d-2})$ are local coordinates on $S$. Among all possible embeddings, the trivial embedding $\phi_0(\sigma)=(0,\sigma^i)$ plays a special role, as it corresponds to bulk coordinates adapted to $S$. In the following, we make a slight abuse of notations and refer to the trivially embedded submanifold simply as $S$ instead of $\phi_0(S)$.
In the region where the $(u^a,x^i)$ coordinates are defined, we can write a vector field in $\mathfrak{X}(M)$ as $\xi=\xi^a\partial_a+\underline{\xi}^i\partial_i$.
Taking the trivial embedding, one can expand both components around points on $S$ as
\begin{align}
\label{Eq.vectorfieldexpansion}
    &\xi^a(u,x)=\xi^a_{(0)}(x)+\xi^a_{(1)b_1}(x)u^{b_1}+\frac{1}{2}\xi^a_{(2)b_1 b_2}(x)u^{b_1}u^{b_2}+\dots,\\
    &\underline{\xi}^i(u,x)=\underline{\xi}^i_{(0)}(x)+\underline{\xi}^i_{(1)b_1}(x)u^{b_1}+\frac{1}{2}\underline{\xi}^i_{(2)b_1 b_2}(x)u^{b_1}u^{b_2}+\dots.
\end{align}
It is easy to see that a consistent truncation in $\{u^1,u^2\}$ has to stop at 0th order for $\xi^i$ and at 1st order for $\xi^a$. Indeed, as soon as we include other terms, the Lie brackets force us to include the entire infinite sequence.
Let us denote these truncated vector fields as
\begin{align}
    &\hat\xi=\underline{\xi}^i_{(0)}(x)\partial_i+(\xi^a_{(0)}(x)+\xi^a_{(1)b}(x)u^{b})\partial_a.
\end{align}
The subalgebra they generate is $\dS\loplus(\Liegl{2}\loplus \RR^2)^S$, as we can see explicitly from their Lie brackets
\begin{equation}
    \begin{split}
        \Lieb{\hat\xi,\hat\xi'}=&\underset{\dS}{\underbrace{\Lieb{\underline{\xi}_{(0)},\underline{\xi'}_{(0)}}}}
        +\underset{\dS\;\text{acts on}\; \RR^2}{\underbrace{\pqty{\underline{\xi}_{(0)}(\xi'^a_{(0)})-\underline{\xi'}_{(0)}(\xi_{(0)}^a)}}}\partial_a
        +\underset{\dS\;\text{acts on}\; \Liegl{2}}{\underbrace{\pqty{\underline{\xi}_{(0)}(\xi'^a_{(0)b})-\underline{\xi'}_{(0)}(\xi_{(0)b}^a)}}}u^b\partial_a\\
        &-\underset{\Liegl{2}}{\underbrace{\lbr\xi_{(1)},\xi'_{(1)}\rbr^b{}_a}}u^b\partial_a
        -\underset{\Liegl{2}\;\text{acts on}\;\RR^2}{\underbrace{\pqty{\xi^a_{(1)b}\xi'^b_{(0)}-\xi'^a_{(1)b}\xi^b_{(0)}}}}\partial_a.
    \end{split}
\end{equation}
Here, we used $\lbr\cdot,\cdot\rbr$ to distinguish the Lie brackets of $\Liegl{2}$ -- the usual matrix commutator -- from the Lie brackets of vector fields. The algebra closes because $\Lieb{\hat\xi,\hat\xi'}$ is another consistently truncated vector field.
This algebra contains the diffeomorphisms of the corner (which are one-to-one with the diffeomorphisms of $S$), and the set of smooth maps $S\to \Liegl{2}\ltimes \RR^2$, that we can think as a copy of the target algebra for each point on the corner. 

Since we did not make any reference to any field theory content or geometry, the algebra we identified is kinematical and universal in the following sense: any theory with diffeomorphism invariance is trivially invariant under the subalgebra, but most of the diffeomorphisms correspond to pure gauge transformations even in the presence of a corner; what is non trivial is that this subalgebra contains the algebra of corner charges of all geometric theories of gravity. By virtue of this universality, we call $\mathfrak{ucs}(S)=\dS\loplus(\Liegl{2}\loplus \RR^2)^S$ the universal corner symmetry (UCS) algebra.~\footnote{It was previously referred to as the \textit{maximal embedding algebra}.} The construction we just performed gives its so-called \textit{defining representation}. Classifying all the (irreducible) unitary representations of the $\mathfrak{ucs}(S)$ gives us control over the possible quantum states describing gravity in a subregion bounded by $S$.

So far, there are no known examples of theories whose charges at a specific corner realize the full $\mathfrak{ucs}$ algebra. Nevertheless, we can find two important subalgebras. The first is the \textit{extended corner symmetry} (ECS) algebra
\begin{equation}
    \mathfrak{ecs}:\quad \dS\loplus (\Liespl{2}\loplus \RR^2)^S
\end{equation}
which is realized at finite-distance corners.\footnote{In General Relativity, given an \textit{isolated} corner (in contrast to a corner that is part of a codimension 2 foliation of spacetime), one can locally write the metric in the Randers-Papapetrou form~\cite{randers_asymmetrical_1941, papapetrou_champs_1966}
\begin{equation*}
    \dd s^2=h_{ab}(u,x)(\dd u^a-\mathrm{a}^a_i(u,x) \dd x^i)(\dd u^b-\mathrm{a}^b_j(u,x)\dd x^j)+\gamma_{ij}(u,x)\dd x^i \dd x^j,
\end{equation*}
and expand each metric function in $\{u^1,u^2\}$ in a similar fashion to~\eqref{Eq.vectorfieldexpansion}.
The pull-back of the diffeomorphism Noether charge aspect $-$ to the corner gives the following generators of the corner symmetry
\begin{equation*}
    H_{\xi}=\xi_{(1)}^a{}_b N^b{}_a+\xi_{(0)}^a p_a +\xi^j_{(0)} b_j,
\end{equation*}
where
\begin{align}
    &N^b{}_a=\sqrt{-h^{(0)}}h_{(0)}^{bc}\epsilon_{ca},\\
    &p_a=\frac{1}{2}\sqrt{-h^{(0)}}\epsilon^{bc}(h^{(1)}_{ab,c}-h^{(1)}_{ac,b}),\\
    &b_i=\sqrt{-h^{(0)}}h_{(0)}^{bc}\epsilon_{ca}\mathrm{a}_i^{a(1)}{}_{b}.
\end{align}
Since $H_\xi=H_{\hat\xi}$, we see that only the vector fields in the ECS subalgebra have non-trivial charges: the determinant subgroup of $\gl{2}$ is inactive because $N^a{}_b$ is traceless.
}
The second, called \textit{asymptotic corner symmetry} (ACS) algebra, is
\begin{equation}
    \mathfrak{acs}: \quad \dS\loplus (\mathfrak{w}\loplus \RR^2)^S,
\end{equation}
where $\mathfrak{w}\cong\RR$ is the subalgebra associated to the determinant in $\gl{2}$. This algebra contains the BMSW algebra, which was shown to be the most general that is realized in asymptotically flat geometries~\cite{freidel_weyl_2021}. All these algebras share the structure $\dS\ltimes \h^S$ (we repeat that $\h^S\equiv C^\infty(S,\h)$), where the local algebra $\h$ is $\h_{\mathrm{G}}:=\Liegl{2}\loplus\RR^2$ for the universal algebra, $\h_{\mathrm{S}}:=\spl{2}\loplus\RR^2$ for the extended, and $\h_{\mathrm{W}}:=\mathfrak w\loplus \RR^2$ for the asymptotic one.
Notice that both $\h_{\mathrm{S}}$ and $\h_{\mathrm{W}}$ are ideals in $\h_{\mathrm{G}}$.

In this work, we will focus on the local algebra $\h$, dropping the $\dS$ part of the group. One can then interpret our results as either stemming from a point-wise analysis on $S$, a restriction to the spherically symmetric sector of phase space, or as a complete analysis in $d=2$, where the corner is just a pair of points.
Moreover, we will only consider the ECS as our classical symmetry group.

In the following, we provide a description of the algebra $\h_S=\Liespl{2}\loplus \RR^2$ and some of its properties. We define the $\mathfrak{sl}(2,\mathbb R)$ algebra as the vector space spanned by $\{\el{L}_0,\el{L_\pm}\}$ with the following Lie brackets
\begin{equation}
    [\el{L}_-,\el{L}_+]=-2i\el{L}_0,\quad [\el{L}_0,\el{L_\pm}]=\mp i \el{L_\pm}.
\end{equation}
This algebra has the well-known quadratic Casimir element, corresponding to the squared momentum~\footnote{As a vector space, $\spl{2}=\RR^{(1,2)}$, and the adjoint action of the algebra on itself corresponds to the set of 2d Lorentz transformations: one rotation and two boosts.}
\begin{equation}
\label{eq.SL2Casimir}
\el{C}_2^{\Liespl{2}}=\el{L}^2\equiv\eta^{ij}\el{L}_i\el{L}_j=-\el{L}_0^2+\frac{1}{2}(\el{L}_+\el{L}_-+\el{L}_-\el{L}_+),
\end{equation}
where we introduced indices $i,j=-,0,+$ and the invariant Killing metric $\eta^{ij}$ with signature $(1,2)$. 
To build $\h_{S}$, we include the generators $\el{T}_\pm\in\RR^2$ (with $[\el{T}_-,\el{T}_+]=0$) in our algebra and extend the Lie brackets to them in the following way
\begin{equation}\label{Eq.ecsalgerba}
    [\el{L}_0,\el{T_\pm}]=\mp \frac{i}{2}\el{T_\pm},\quad [\el{L_\pm},\el{T_\pm}]=0,\quad [\el{L_\pm},\el{T_\mp}]=\pm i\el{T_\pm}.
\end{equation}
In this larger algebra, $C_2^{\Liespl{2}}$ is not a Casimir anymore. Nevertheless, the $\mh_S$ algebra still admits a Casimir, which is cubic in the generators~\cite{ciambelli_universal_2023}
\begin{equation}
\label{eq.HSCasimir}
    \el{C}^{\h_\mathrm{S}}_3=-\el{L}_+ \el{T}_-^2-\el{L}_- \el{T}_+^2+\pqty{\el{L}_0+\frac{3}{8}i} \el{T}_+\el{T}_-+\pqty{\el{L}_0-\frac{3}{8}i}\el{T}_-\el{T}_+.
\end{equation}
At this stage, one could simply drop the $3/8$ factors because $\el{T}_-$ and $\el{T}_+$ commute, but we prefer to keep them around for later convenience.

\subsection{Quantum extension}

When we consider a quantum theory, we are interested in the construction of a Hilbert space $\mathcal H$ and the computation of transition amplitudes $\braket{a}{b}$ between any two (normalized) vectors $a,b\in\mathcal H \backslash \{0\}$. On the other hand, since we can only ever have access to transition probabilities $\abs{\braket{a}{b}}^2$, we cannot determine the phases of each vector, but only their difference. For this reason, quantum states are defined as equivalence classes of normalized vectors $\ket{a}\sim e^{i\theta}\ket{a}$ (i.e. they are element of the projective Hilber space $(\mathcal H \backslash \{0\})/\CC^\times$), and the operators $U_g$ that represent a symmetry transformation $g$ are allowed to change $\theta$: Since this change might depend on the transformation itself, one finds that they generically form a \textit{projective} representation of the group. This means that two operators compose as
\begin{equation}
    U_{g_1} U_{g_2}=\omega(g_1,g_2)U_{g_1 g_2},
\end{equation}
where $\omega(g_1,g_2)$ is a 2-cocycle of the symmetry group. An important result in representation theory is that the projective representations of a path-connected group are equivalent to the ordinary representations of its maximal central extension~\cite{bargmann_unitary_1954, mackey_unitary_1958}.

In light of these considerations, we expect that the symmetry group which is going to be represented in the Hilbert space is not $\spl{2}\ltimes\RR^2$, but rather its maximal central extension, which we dub \textit{quantum corner symmetry} (QCS) group and denote by
\begin{equation}
\label{Eq.HQgroup}    H_Q=\widetilde{\spl{2}}\ltimes\HH_3
\end{equation}
where the tilde denotes the universal cover and $\HH_3$ is the Heisenberg group of 1d quantum mechanics, that is a 3 dimensional Lie group. The extension $\RR^2\to \HH_3$ is due to the presence of a non-trivial 2-cocycle in the translation sector of the ECS, whereas the universal cover is due to the fact that $\spl{2}$ is not simply connected. Whereas the latter extension is present at the classical level -- a group and its universal cover share the same algebra -- the former is a purely quantum effect, analogous to the appearance of a non-trivial commutator between positions and momenta in quantum mechanics.

Let us now consider the Lie algebra associated to $H_Q$, which we can also think as a central algebra extension of $\h_S$. To construct $\h_\mathrm{Q}$, we introduce a central element in the translation sector
\begin{equation}
    [\el{T}_-,\el{T}_+]=-i\el E,
\end{equation}
making it non-abelian $\RR^2\to \HH_3$.
Because of this deformation, the cubic Casimir is deformed as well
\begin{equation}
\label{eq.HQCasimir}
    \el{C}^{\h_\mathrm{Q}}_3=\el{C}^{\h_S}_3+2\el E \,\el{C}_2^{\Liespl{2}}
\end{equation}
Clearly, in the limit where the deformation is removed ($\el{E}\to 0$), we recover the Casimir of $\h_S$. What is more surprisingly is that, in the opposite limit $\el E\to \infty$, $\el{C}_3^{\h_\mathrm{Q}}$ becomes proportional to $\el{C}_2^{\Liespl{2}}$.
This apparent coincidence hints at a quite remarkable property of this algebra, which we are now going to exhibit. Let us consider the universal enveloping algebra of $\h_\mathrm{Q}$ and define the following elements~\footnote{The universal enveloping algebra does not contain the reciprocal of $\el E$. Strictly speaking, the $\elg{\DL}$ generators belong to a quotient of the universal enveloping algebra of $\RR\times\h_\mathrm{Q}$, where the basis element $\el{E}^{-1}\in \RR$ commutes with everything, by the ideal generated by $\el E^{-1} \el E-1$ (or, equivalently, $\el E \el E^{-1}-1$).}
\begin{subequations}
\label{Eq.Mgenerators}
\begin{gather}
    \elg{\DL}_0=\el{L}_0-\frac{1}{4\el E}(\el{T}_+ \el{T}_-+\el{T}_- \el{T}_+),\\
    \elg{\DL}_+=\el{L}_+-\frac{1}{2\el E}\el{T}_+^2,\qquad \elg{\DL}_-=\el{L}_--
    \frac{1}{2\el E}\el{T}_-^2,
\end{gather}
\end{subequations}
which satisfy an $\Liespl{2}$ algebra
\begin{equation}
    [\elg{\DL}_-,\elg{\DL}_+]=-2i \elg{\DL}_0, \quad [\elg{\DL}_0, \elg{\DL}_\pm]=\mp i  \elg{\DL}_\pm,
\end{equation}
and commute with the translations
\begin{equation}
\label{Eq.AlegbraFactorization}
    [\elg{\DL}_0,\el{T}_\pm]=0,\quad [\elg{\DL}_\pm,\el{T}_\pm]=0,\quad [\elg{\DL}_\pm,\el{T}_\mp]=0.
\end{equation}
This construction implies that we can embed the algebra of $\spl{2}\times\HH_3$ in the universal enveloping of the algebra of $\spl{2}\ltimes\HH_3$. In terms of these modified generators, the cubic Casimir of $\h_\mathrm{Q}$ can be written as 
\begin{equation}
\label{Eq.QCScasimirM}
    \el{C}_3^{\h_\mathrm{Q}}=2\el{E}\pqty{-\elg{\DL}_0^2+\frac{1}{2}(\elg{\DL}_+ \elg{\DL}_-+\elg{\DL}_- \elg{\DL}_+)},
\end{equation}
reproducing the $\spl{2}$ invariant product $\elg{\DL}^2$. The limit $\el{E}\to\infty$ is the limit in which the quadratic $\el{T}$ contributions in~\eqref{Eq.Mgenerators} becomes negligible and the new $\Liespl{2}$ generators approach the old one $\elg{\DL}\to \el{L}$. Accordingly, $\el{C}_3^{\h_\mathrm{Q}}$ tends to $2\el{E}\el{L}^2$.
One can also rewrite the cubic $\h_S$ Casimir in the intriguing form
\begin{equation}
    \el{C}_3^{\h_S}=2\el{E}(\elg{\DL}^2-\el{L}^2).
\end{equation}
Let us now give an interpretation of the elements~\eqref{Eq.Mgenerators}. Since both $\el{L}$ and $\elg{\DL}$ generate a $\Liespl{2}$ algebra, it follows that
\begin{equation}
\label{Eq.Metaplecticgenerators}
    \frac{1}{4\el{E}}(\el{T}_+ \el{T}_-+\el{T}_- \el{T}_+),\qquad 
    \frac{1}{2\el{E}}\el{T_+}^2,\qquad 
    \frac{1}{2\el{E}}\el{T_-}^2,
\end{equation}
are $\Liespl{2}$ generators as well. These elements are well-known in the mathematical literature as the generators of the \textit{Weil representation}, which often arises in the context of induced representations.
Since $\spl{2}$ is the automorphism group of $\HH_3$, representing $\el{T}_+$ and $\el{T}_-$ in some space automatically induces an $\mathrm{Mp}(2,\mathbb{R})$ representation in the same space, where $\mathrm{Mp}(2,\mathbb{R})$ is the \textit{metaplectic group}, the double cover of $\spl{2}$. In this sense sense, the $\elg{\DL}$ generators parametrize the "distance" of a certain $\spl{2}$ representation from the metaplectic one.

Let us take a moment to pause and consider what happens in $d>2$. If we were to look back at the entire corner, given the common structure of the universal corner symmetry algebras, one could naively expect the full quantum group to be something like $\widetilde{\text{Diff}(S)}\ltimes H_\mathrm{Q}$, where $\widetilde{\text{Diff}(S)}$ is the maximal central extension of $\text{Diff}(S)$. However -- in addition to the complication that the maximal central extension of the diffeomorphisms group is highly dependent on the dimension and the topology of $S$ -- the central extension of a (semidirect) product is not the (semidirect) product of the extensions, which means that finding the correct group requires looking for the extensions of the ECS group $\text{Diff}(S)\ltimes(\spl{2}\ltimes\RR^2)^S$ itself. Already at the classical level, we know that reintroducing diffeomorphisms has dramatic consequences. At the algebra level, $\dS\loplus \h_\mathrm{S}$ and $\dS\loplus \h_\mathrm{W}$ are no longer ideals inside $\dS\loplus \h_\mathrm{G}$. Diffeomorphisms can change both the form and the number of Casimirs. For example, in the simpler case when we drop the translation sector (what remains are \textit{corner-preserving} transformations), we know that $\Liespl{2}$ has a single independent Casimir, while $\dS\loplus \Liespl{2}^S$ has an infinite number of invariants~\cite{donnelly_gravitational_2021}.

\subsection{Representation theory of the quantum corner symmetry group}

To keep the exposition self-contained, we report here the complete set of unitary irreducible representations of $H_Q$ that was found in~\cite{varrin_physical_2024}, together with a presentation of the algebraic method one can use to find them.

Let us consider a semidirect product group $H\ltimes N$. By this, we mean that elements of $H$ are associated to automorphisms of $N$ ($n\mapsto h\lga n$) such that the composition law has a mixing between the two sectors
\begin{equation}
    (h_1, n_1)\cdot (h_2, n_2)=(h_1\cdot h_2, n_1\cdot(h_1\lga n_2)).
\end{equation}
If $\pi$ is a irreducible representation of $N$ on the target space $V_\pi$, then $h\in H$ acts on it producing a new representation $\pi_h$ in the same space, defined by 
\begin{equation}
    (\pi_h)(n)=\pi(h^{-1}\lga n).
\end{equation}
Considering the equivalence class of both sides (two representations are equivalent if there is an isomoprhism $\mathcal I:V_{\pi_1}\to V_{\pi_2}$ such that $\pi_2 \circ \mathcal I= \mathcal I\circ \pi_1$), one extends the action of $H$ to the set of inequivalent irreducible representations of $N$ ($[\pi]_h=[\pi_h]$). This set is foliated by $H$-orbits. 

Some classes of inequivalent representations have non-trivial stabilizers, which means that there exists some elements of $H$ that leave the class invariant. These elements form a subgroup that we call the \textit{little group} of that class
\begin{equation}
    H_{[\pi]}=\{h\in H| [\pi_h]=[\pi]\}
\end{equation}
This definition means that, for each $\pi\in [\pi]$, there exists an automorphism $\mathcal I_h\in \Aut(V_\pi)$ such that $\pi_h=\mathcal I_h \circ \pi \circ \mathcal I_h^{-1}$.

According to Mackey's theory of induced representations~\cite{mackey_unitary_1958,mackey_theory_1976}, one can construct the irreducible (unitary) representations of the group $H\ltimes N$ by first constructing the irreducible (unitary) representations of $H_{[\pi]}\ltimes N$, and then inducing a representation of the full group~\footnote{Given a group $G$ and a representation $D:H\to \End(V)$ of a subgroup of $H\subset G$, one can construct a representation of the $G$ over the space of $V$-valued functions of the coset $G/H$ by $g\mapsto U_g\in \End(\Gamma(G/H,V))$ with
\begin{equation*}
    (U_{g}\psi)[p]=D(h)\,\psi[g^{-1}\lga p],\quad\forall p\in G/H
\end{equation*}
where $\lga$ denotes here the natural action of $G$ on the coset $G/H$.}. In the case where $H_{[\pi]}$ coincides with the full group, one does not need to perform the last step and the representations take the form of a tensor product {($\End(V,W)$ is the space of linear maps $V\to W$, here $\End(V)\equiv\End(V,V)$)
\begin{align}
    \rho_{\sigma,[\pi]}:H_{[\pi]}&\ltimes N\to \End(\mathcal H_\sigma\otimes V_\pi),\\
    (h&,n)\mapsto \rho_{\sigma,[\pi]}(h,n)=\sigma(h)\otimes \pi(n)\mathcal I_h,
\end{align}
where $\sigma: H_{[\pi]}\to \End(\mathcal H_\sigma)$ is a representation of $H_{[\pi]}$. The complication when $N$ is non abelian is that the maps $h\to \mathcal I_h$ generally form a projective (unitary) representation of $H_{[\pi]}$. Therefore, before combining the representations of $H_{[\pi]}$ and $N$, $\sigma$ needs to a projective representation that "balances" $\mathcal I$ in the sense that $h\to \sigma(h)\otimes  \mathcal I_h$ is an ordinary, non projective, representation.

In the case at hand, the QCS group $H_Q$ in~\eqref{Eq.HQgroup} has the structure of a semidirect product with a non-abelian normal subgroup, $\mathbb H_3$.
According to Stone-Von Neumann's theorem~\cite{stonevonneumann}, the Heisenberg group has essentially one, up to equivalence, (strongly continuous) unitary irreducible representation $\pi_e$ for each choice of non-zero $i e=\pi^*_e(\el{E})\in i\RR$ -- an element in the center $Z(\h_3)$ must be represented by a multiple of the identity in any irreducible representation. For $e=0$, there are infinite irreducible representations, but they are not interesting for us. Therefore $\hat\HH_3=\RR$. One possible representation space for $\pi_e$ is the Fock space $\Fock_e$ of a free particle with ladder operators
\begin{subequations}
    \label{Eq.ladder}
\begin{gather}
    \el{a}=\frac{1}{\sqrt{e}}\pi^*_e(\el{T_-}),\quad \el{a}^\dagger=\frac{1}{\sqrt{e}}\pi^*_e(\el{T_+}), \quad{\text{if}}\quad e>0,\\
    \el{a}=\frac{1}{\sqrt{\abs{e}}}\pi^*_e(\el{T_+}),\quad \el{a}^\dagger=\frac{1}{\sqrt{\abs{e}}}\pi^*_e(\el{T_-}), \quad{\text{if}}\quad e<0.
\end{gather}
\end{subequations}
The little group of $\pi_e$ is the entire $\widetilde{\spl{2}}$, which means that the representation theory of $H_Q$ falls in the simple case where the representations are a tensor product.

The representations of $\widetilde{\spl{2}}$ depend on two numbers, associated to the eigenvalue of the Casimir $\el{L}^2$ and to the eigenvalue of the central element $e^{2\pi \el{L_0}}$~\cite{kitaev_notes_2018}. We write the first as $q(1-q)$, with $q\in\RR$ or $q-\frac{1}{2}\in i\RR$, and the second as $e^{2\pi i\mu}$. When the group that is represented is the non-covered $\spl{2}$, the central element must be in $\ZZ_2$, which means that $\mu\in \ZZ/2$. If we call $\mathcal H_{q,\mu}$ the carrying space of this representation, we deduce that representations of the QCS group are $\rho_{q,\mu,e}\equiv\rho_{\sigma_{q,\mu},[\pi_e]}:H_Q\to\End(\mathcal H_{q,\mu}\otimes \Fock_e)$.
At the algebra level, the basis elements of $\h_Q$ are represented as~\footnote{This for $e>0$. The inequivalent representation for $e<0$ can be easily constructed.}
\begin{subequations}
\label{Eq.VarrinReps}
    \begin{align}
        &\rho^*_{q,\mu,e}(\el{T}_-)\ket{n,k}=\sqrt{e}\sqrt{k}\ket{n,k-1},\\
        &\rho^*_{q,\mu,e}(\el{T}_+)\ket{n,k}=\sqrt{e}\sqrt{k+1}\ket{n,k+1},\\
        \label{eq.n def}
        &\rho^*_{q,\mu,e}(\el{L}_0)\ket{n,k}=\pqty{\frac{2k+1}{4}+n}\ket{n,k},\\
        &\rho^*_{q,\mu,e}(\el{L}_-)\ket{n,k}=-\sqrt{(n-q)(n-1+q)}\ket{n-1,k}+\frac{1}{2}\sqrt{k(k-1)}\ket{n,k-2},\\
        &\rho^*_{q,\mu,e}(\el{L}_+)\ket{n,k}=-\sqrt{(n+q)(n+1-q)}\ket{n+1,k}+\frac{1}{2}\sqrt{(k+1)(k+2)}\ket{n,k+2},
    \end{align}
\end{subequations}
where we write $\ket{n,k}$ for $\ket{n}\otimes\ket{k}\in \mathcal H_{q,\mu}\otimes \Fock_e$. {The quantum numbers $k$ is the eigenvalue of the Fock space number operator, while $n$ is defined by~\eqref{eq.n def}.
The cubic Casimir is given by a multiple of the identity
\begin{equation}
    \rho^*_{q,\mu,e}(\el {C}_3^{\h_Q})\ket{n,k}=2e q(1-q)\ket{n,k}.
\end{equation}
The representation in $\Fock_e$ is unitary for all $e\ne 0$, while the special linear operators are unitarily realized in $\mathcal H_{q,\mu}$ only in the following cases:
\begin{description}
    \item[Trivial representation:] When $q=0$ and the spectrum of $\el{L}_0$ contains zero, then the representation is one-dimensional -- it is the only finite-dimensional unitary representation -- because all generators annihilate the state $\ket{0}$.
    \item[Discrete series representation] For $q>\frac{1}{2}$ and $\mu=\pm q$, then $\el{L}_\mp\ket{\pm q}=0$ and the spectrum of $\el{L}_0$ is either $q+\ZZ$ or $-q-
    \ZZ$.
    \item[Continuous series representations:]
    If $q=\frac{1}{2}+i\RR$ and
    $\abs{\mu}\le\frac{1}{2}$ or $\abs{\mu}<q<\frac{1}{2}$ -- these two cases are known as the \textit{principal} and the \textit{complementary} series -- the ladder created by $\el{L}_\pm$ does not stop and the spectrum of $\el{L}_0$ is $\mu+\ZZ$. The representations for $\mu>0$ and $\mu<0$ are equivalent.
\end{description}

\section{The orbit method}
\label{Sec.Method}

\subsection{Review of coadjoint orbits}

Let $G$ be a Lie group and $\g$ its Lie algebra $\g\cong T_e G$. $G$ has a natural action on itself by conjugation $h \mapsto g h g^{-1}$. The push-forward of this map allows us to construct an action of $G$ on its algebra
\begin{equation}
    G\times\g\to\g,\quad (g,X)\mapsto\dv{}{t}\pqty{g e^{t X} g^{-1}}\eval_{t=0}=:\Ad{g}X
\end{equation}

Since the Lie algebra $\g$ is a vector space, we can introduce the dual space $\g^*$ as the space of linear functionals $\el{m}:\g\to\RR$, $\el{X} \mapsto\pair{\el{m}}{\el{X}}\in\RR$.
In this space, we can define the coadjoint representations by requiring consistency with the pairing $\pair{\cdot}{\cdot}$, that is
\begin{equation}
    \cAd{}:G\to \mathrm{GL}(\g^*), \quad \pair{\cAd{g}\el{m}}{\el{X}}=\pair{\el{m}}{\Ad{g^{-1}}\el{X}}.
\end{equation}

Given a point $\el{m}\in\g^*$, we introduce its coadjoint orbit as the set of points that can be obtained acting on it with the group
\begin{equation}
    \mathcal O_{\el{m}}=\Bqty{\cAd{g}\el{m}\,|\,g\in G}.
\end{equation}
It is sometimes useful to add a $G$ superscript to distinguish between coadjoint orbits of different groups.
It is easy to see that coadjoint orbits do not intersect, hence they foliate $\g^*$. Since the action of $G$ is generically not free, the leaves of this foliation do not have the same dimensions. To make this statement precise, we define the \textit{stabilizer} of $\el{m}\in \g^*$ as the subgroup of $G$ elements that act trivially on it
\begin{equation}
    G_{\el{m}}=\Bqty{g\in G\,|\,\cAd{g}\el{m}=\el{m}}.
\end{equation}
An important result in the theory of Lie groups is that the orbits can be written as homogeneous spaces
\begin{equation}
\label{Eq.orbitshomo}
    \mathcal O_{\el{m}}\simeq G/G_{\el{m}}.
\end{equation}

The relevance of the coadjoint representation is that the corresponding orbits are symplectic manifolds. In fact, $\g^*$ inherits a Poisson algebra structure from the Lie algebra structure of $\g$
\begin{equation}
\label{Eq.KKPoissonBrackets}
    \{\mathcal X,\mathcal Y\}(\el{m})=\pair{\el{m}}{[\mathcal X_{*\el{m}},\mathcal Y_{*\el{m}}]},\quad \forall \mathcal X,\mathcal Y\in C^{\infty}(\g^*)
\end{equation}
where the differential $\mathcal X_{*\el{m}}$ of any function at $\el{m}$ is a linear map $T_{\el{m}}\g^*\to T_{\mathcal X(\el{m})}\RR$. Since both $\g^*$ and $\RR$ are linear spaces, this is also a map $\g^*\to\RR$, hence an element of $(\g^*)^*$, which we can identify (at least for finite dimensional vector spaces) with $\g$, thus explaining why we can take a Lie bracket between two such elements.\footnote{\label{fn.AlgebraVectorFieldsCorrespondence}The converse --- that, for every element $\el{X} \in \g$ and $\el{m}\in\g^*$, one can find a function $\mathcal X$ such that $\mathcal X_{*\el{m}}=\el{X}$--- is always true.}

With the Poisson brackets at our disposal, we can associate to any function $\mathcal X\in C^{\infty}(\g^*)$ its \textit{Hamiltonian vector field} $\xi_\mathcal X\in\mathfrak{X}(\g^*)$
\begin{equation}
    \xi_\mathcal X:=-\{\mathcal X,\cdot\}.
\end{equation}
Since
\begin{equation}
\label{Eq.vfrepresentation}
    [\xi_{\mathcal X},\xi_{\mathcal Y}]_{\mathrm{Lie}}=-\xi_{\{\mathcal X,\mathcal Y\}},
\end{equation}
the distribution of Hamiltonian vector fields is integrable and, by Frobenius' theorem, Poisson manifolds are foliated by their orbits. This foliation is called \textit{symplectic} because each leave is naturally endowed with the closed, non-degenerate 2-form $\omega(\xi_{\mathcal X},\xi_{\mathcal Y})=\{\mathcal X,\mathcal Y\}$.

We finally come to the result that coadjoint orbits are symplectic manifolds by showing that the orbit foliation is equivalent to the symplectic foliation.
From~\eqref{Eq.KKPoissonBrackets} we show that, for each $\mathcal Y$
\begin{equation}
    (\xi_\mathcal X)_{\el{m}}[\mathcal Y]=-\{\mathcal X,\mathcal Y\}(\el{m})=-\pair{\el{m}}{[\mathcal X_{*\el{m}},\mathcal Y_{*\el{m}}]}=-\pair{\el{m}}{\ad{\mathcal X_{*\el{m}}}\mathcal Y_{*\el{m}}]}=\pair{\cad{\mathcal X_{*\el{m}}}\el{m}}{\mathcal Y_{*\el{m}}}.
\end{equation}
Therefore, given a basis $\{\el{e}_i\}$ in $\g$, we can write this vector field as
\begin{equation}
\label{Eq.KKEquivalence}
    (\xi_\mathcal X)_{\el{m}}=\pair{\cad{X}\el{m}}{\el{e}_i}\pdv{}{u_i}\eval_{\el{m}}=:(\xi_{\el{X}})_{\el{m}},
\end{equation}
where the coordinates $\{u_i\}$ are, at each point $\el{q}\in\mathcal O_{\el{m}}$, the components of the same point $\el{q}$ in the basis of $\g^*$ dual to $\{\el{e}_i\}$, and we denoted by $\el{X}$ the Lie algebra element associated to $\mathcal X_{*\el{m}}$.
This explicit construction shows that the tangent space of a symplectic leave is spanned by the orbit generators at each point, whence it follows (by integrability) that the symplectic leave is the orbit itself.
The symplectic structure on each coadjoint orbit is given by the so-called Kirrilov-Kostant-Souriau (KKS) 2-form $\Omega\in\Omega^2(\g^*)$, defined, for each $\el{q}\in\mathcal O_{\el{m}}$, by
\begin{equation}
    \Omega^G_{\el{q}}(\xi_{\el{X}},\xi_{\el{Y}})=\pair{\el{q}}{[\el{X},\el{Y}]}.
\end{equation}

If we denote the infinitesimal generator of the coadjoint action as $\xi_{\el{X}}\sim\cad{\el{X}}$, we can derive the relationship
\begin{equation}
\label{Eq.KKIntegrability}
    I_{\xi_{\el{X}}}\Omega\equiv \Omega(\xi_{\el{X}},\cdot)=\bbd \mommap_{\el{X}}
\end{equation}
where $\bbd$ is the usual exterior derivative on the exterior algebra $\Omega(\mathcal O_{\el{m}})$ and $\mommap_{\el{X}}\equiv \pair{\mommap(\cdot)}{X}$, with $\mommap:\mathcal O_{\el{m}}\to \g^*$ is the so-called \textit{momentum map},\footnote{For a general symplectic manifold $(\mathcal M,\Omega)$, a momentum map is a smooth map $\mommap:\mathcal M\to \g^*$ such that
\begin{equation*}
    I_{\xi_{\el{X}}}\Omega=\delta\pair{\mommap(\cdot)}{\el{X}},\quad\forall \el{X}\in\g.
\end{equation*}}
given by the inclusion $\mathcal O_{\el{m}} \hookrightarrow \g^*$: $\mommap_{\el{X}}(\cAd{g}\el{q})=\pair{\mommap(\cAd{g}\el{q})}{\el{X}}=\pair{\cAd{g}\el{q}}{\el{X}}$.
Indeed $(I_{\xi_{\el{Y}}}\delta\mommap_{\el{X}})_{\el{q}}=(\xi_{\el{Y}}[\mommap_{\el{X}}])_{\el{q}}=\dv{}{t}\pqty{\mommap_{\el{X}}(\cAd{e^{t\el{Y}}}\el{q})}|_{t=0}=\mommap_{\el{X}}(\cad{\el{Y}}\el{q})=\pair{\cad{\el{Y}}\el{q}}{X}=\pair{\el{q}}{[\el{X},\el{Y}]}$, which is the same as $(I_{\xi_{\el{Y}}}I_{\xi_{\el{X}}}\Omega)_{\el{q}}$ for any $\el{q}\in\mathcal O_{\el{m}}$ and any $Y\in\g$.

An important consequence of the existence of the momentum map is that the $G$ action on the orbit is always symplectic, in the sense that each coadjoint transformation is a symplectomorphism
\begin{equation}
    \Lie_{\xi_{\el{X}}}\Omega=\bbd I_{\xi_{\el{X}}}\Omega+\overset{\bbd\Omega=0}{\overbrace{I_{\xi_{\el{X}}}\bbd\Omega}}=\bbd^2 \mathcal \mommap_{\el{X}}=0.
\end{equation}
The momentum map also provide an explicit realization of a function whose differential equals $\el{X}$ at $\el{q}\in\mathcal O_{\el{m}}$, that is $(\mommap_{\el{X}})_{*\el{q}}=\el{X}$. Therefore, the Hamiltonian vector field associated with the momentum map is the generator itself
\begin{equation}
    (\xi_{\mommap_{\el{X}}})_{\el{q}}=(\xi_{\el{X}})_{\el{q}},\quad \forall \el{q}\in \mathcal O_{\el{m}}.
\end{equation}
As a consequence, the momentum map provides a representation of the Lie algebra\footnote{For a generic phase space which is not a coadjoint orbit, the algebra is represented up to central extension.}
\begin{equation}
    \{\mommap_{\el{X}},\mommap_{\el{Y}}\}=\mommap_{[\el{X},\el{Y}]}.
\end{equation}

\paragraph{Orbits of the Heisenberg group}
\label{Par.HeisenbergOrbits}

The Heisenberg group $\HH_3$ can be realized as the set of $3\times3$ upper triangular matrices of the form
\begin{equation}
\label{Eq.defrepHeisenberg}
    n(a,b,c) = \mqty(1&a&c\\0&1&b\\0&0&1),
\end{equation}
with standard matrix multiplication as the group product.
The Heisenberg algebra $\mathfrak{h}_3$ is then isomorphic to the algebra of strictly upper triangular matrices with the standard matrix commutator. A convenient basis is
\begin{equation}
    \el{A} = \mqty(0&1&0\\0&0&0\\0&0&0), \quad \el{B} = \mqty(0&0&0\\0&0&1\\0&0&0), \quad \el E=\mqty(0&0&1\\0&0&0\\0&0&0),
\end{equation}
where we introduce $\el{A}$ and $\el{B}$ by
\begin{equation}
\label{eq.HeisenbergRotation}
    \el{T}_\pm=\frac{\el{A}\pm i \el{B}}{\sqrt{2}},
\end{equation}
so that they satisfy
\begin{equation}
    [\el{A},\el{B}]=\el{E}.
\end{equation}
We now want to describe the elements of $\mathfrak{h}_3^*$ using the standard bilinear form on matrices
\begin{equation}
    \pair{\el p}{\el X} = \Tr[\el p \el X], \quad \el X\in\h,\,\el p \in \h^*.
\end{equation}
Using the above form, we observe that the orthogonal space to the algebra $\h_3^\perp$ is isomorphic to the upper triangular matrices. The dual algebra is therefore isomorphic to strictly lower triangular matrices
\begin{equation}\label{Eq.dualalgebraisomorphismwithmatrices}
    \h_3^* \cong \mathrm{M}_{3\times3}(\mathbb{R})/ \h_3^\perp,
\end{equation}
and is spanned by
\begin{equation}
   \elg{\alpha} = \mqty(0&0&0\\1&0&0\\0&0&0), \quad \elg{\beta}=\mqty(0&0&0\\0&0&0\\0&1&0), \quad \elg{\varepsilon} = \mqty(0&0&0\\0&0&0\\1&0&0).
\end{equation}
Equation \eqref{Eq.dualalgebraisomorphismwithmatrices} defines an equivalence relation where two matrices are equivalent if they have the same strictly lower triangular part. Any matrix $\el p \in \mathrm{M}_{3\times3}\qty(\mathbb{R})$ can then be identified to a element of the dual algebra by considering its equivalence class $[\el{p}]\in\h_3^*$. The coadjoint action of a group element is then simply given by the matrix product
\begin{equation}
   \cAd{n} [\el p] =\qty[n \el p n^{-1}].
\end{equation}
A simple calculation gives
\begin{equation}
   \cAd{n(a,b,c)}\qty[\mqty(0&0&0\\
   A&0&0\\E&B&0)] = \qty[\mqty(0&0&0\\A+b E&0&0\\E&B - a E&0)].
\end{equation}
The fact that $E$ is constants under the coadjoint action arises from the fact that $\el{E}$ is a Casimir of the Heisenberg algebra. At the dual algebra level, a coordinate function associated with a Casimir element is a Casimir function. Since the orbits are characterized by a fixed value of $E$, points $\el{p}=(0,0,E)$ in the coadjoint space with different $E$ belong to separate orbits, and for every one of them we can take one such point as a representative. The above equation tells us that, given a value of $E\neq 0$\footnote{The $E=0$ orbits are simply points and will not be of interest to us.}, the orbits of the Heisenberg group are two-dimensional planes
\begin{equation}\label{Eq.heisenbergorbits}
     \mathcal{O}^{\HH_3}_{E}\equiv \mathcal{O}^{\HH_3}_{(0,0,E)} = \qty{(A,B,E)\mid A,B\in \mathbb{R}}.
\end{equation}
As expected, we have a foliation of the dual algebra in terms of coadjoint orbits
\begin{equation}
    \h_3^* \cong \mathbb{R}^3 = \bigcup_{E\in\mathbb{R}} \mathcal{O}^{\HH_3}_{E}.
\end{equation}

Using the coadjoint action, we can compute the fundamental vector fields $\mathrm{ad}^*_{\el X}$ in the basis of $T\mathcal{O}_{E}^{\HH_3}$ associated with the $(A,B)$ coordinates on the orbit
\begin{equation}
    \xi_{\el{A}} = -E \partial_{B}, \quad \xi_{\el{B}} = E \partial_{A}, \quad \xi_{\el{E}} = 0.
\end{equation}
The only non-vanishing component of the KKS symplectic form at a point $\el{p}=(A,B,E)$ is given by
\begin{equation}
\Omega^{\HH_3}_{\el{p}}\qty(\xi_{\el{A}},\xi_{\el{B}}) =\pair{\el{p}}{[\el{A},\el{B}]}=\pair{(A,B,E)}{\el{E}} = E,
\end{equation}
which implies
\begin{equation}
\label{Eq.H3KKS}
    \Omega^{\HH_3} = \frac{1}{E}\bbd A\wedge \bbd B.
\end{equation}
Since the orbits are contractible -- they are isomorphic to $\RR^2$ -- the KKS form is not only closed but exact. Indeed we can write $\Omega^{\HH_3} = - \bbd \theta^{\HH_3}$ with $\theta^{\HH_3}=\frac{1}{E}B\bbd A$.

\paragraph{Orbits of $\spl{2}$}
\label{Par.SLorbits}

The special linear group is the set of $2\times 2$ matrices of unit determinant with matrix multiplication as the group product. At the algebra level, this condition reflects in the matrix defining representation being traceless. We choose a basis to be
\begin{equation}
    \el{L}_0=\frac{i}{2}\mqty(1 & 0\\ 0& -1), \quad \el{L}_+=\mqty(0 & 0\\ i& 0), \quad \el{L}_-=\mqty(0 & -i\\ 0& 0).
\end{equation}
As in the previous example, we want the pairing between the algebra and its dual to be given by the bilinear form on $\mathrm{M}_{2\times 2}(\RR)$. This implies that $\Liespl{2}^\perp$ are multiples of the identity and that $\Liespl{2}^*$ are traceless matrices as well. Alternatively, one could infer that $\Liespl{2}^*\cong \Liespl{2}$ from the existence of the non-degenerate Killing form $\eta$: taking $\elg{\theta}^{i}=2\eta^{ij}\el{L}_j$ as the basis for the dual space explicitly shows the isomorphism between the two.
Given this basis, the explicit coadjoint action transform these components as
\begin{equation}
    \cAd{h}(L_i\elg{\theta}^i)=2L_j \eta^{jk}\Tr[h \el{L}^k h^{-1} \el{L}_i]\elg{\theta}^i\equiv L_j \Lambda(h)^j{}_i \elg{\theta}^i
\end{equation}
where $\Lambda(h)^j{}_k$ is a $\so{2,1}$ Lorentz transformation. Therefore, the coadjoint orbits of $\spl{2}$ are in one-to-one correspondence with the orbits of a vector in $\RR^{(1,2)}$ -- given by the coordinates $(L_0,L_+,L_-)$ -- under Lorentz transformations, from which we borrow the nomenclature. These orbits are characterized by the constant value of $L^2\equiv -\Delta$, which is the Casimir function associated to $\el{L}^2$:
\begin{itemize}
    \item Massive orbits: When $\Delta>0$, the orbits are the two sheets of a two-sheeted hyperboloid, distinguished by the sign of $L_0$.
    The corresponding tangent spaces are spanned by~\footnote{The coadjoint orbit generators $\xi_{\el{L}_i}$ can be written in the coordinate basis of $\Liespl{2}^*$ as
        \begin{equation*}
       \xi_{\el{L}_0}=iL_+\partial_{L_+}-iL_-\partial_{L_-},\quad\xi_{\el{L}_+}=-iL_+\partial_{L_0}-2iL_0\partial_{L_-},\quad\xi_{\el{L}_-}=iL_-\partial_{L_0}+2iL_0\partial_{L_+}.
        \end{equation*}}
    \begin{equation}
        \partial_{L_+}=\mp\frac{i}{2\sqrt{\Delta+L_+ L_-}}\xi_{\el{L}_-},\quad \partial_{L_-}=\pm\frac{i}{2\sqrt{\Delta+L_+ L_-}}\xi_{\el{L}_+},
    \end{equation}
    where the upper (lower) sign corresponds to the upper (lower) sheet.
    On these orbits, the KKS form is
    \begin{equation}
    \Omega_{\Delta,\pm}^{\spl{2}}=\pm\frac{i}{2}\frac{\bbd L_-\wedge \bbd L_+}{\sqrt{\Delta+L_+ L_-}}.
    \end{equation}
    \item Tachyonic orbits: When $\Delta<0$, the orbits are one-sheeted hyperboloids. To conveniently parametrize them we use the angle $\varphi$, defined by $L_\pm=(L_0\pm i s)\,e^{\mp i\varphi}$ with $s^2=-\Delta$, which allows to span the tangent space to these orbits with
    \begin{equation}
        \partial_{L_0}=\frac{i}{2}\pqty{\frac{L_0+is}{L_0}\frac{\xi_{\el{L}_+}}{L_+}-\frac{L_0-is}{L_0}\frac{\xi_{\el{L}_-}}{L_-}},\quad \partial_\varphi=-\xi_{\el{L}_0}.
    \end{equation}
    Computing the KKS form gives
    \begin{equation}
    \label{Eq.tachyonicKKS}
    \Omega_{\Delta}^{\spl{2}}=\bbd L_0\wedge \bbd \varphi.
    \end{equation}
    \item Trivial orbits: When $\Delta=0$ because $L_i=0$, the orbit is a single point.
    \item Null orbits: When $\Delta=0$ but $L_i\ne 0$, there are only two orbits,
    given by the null cones in the $L_0>0$ and $L_0<0$ half-spaces. One can see the null orbits either as the $\Delta\to 0^+$ limit of the punctured massive orbits (one removes the point $L_
    +=L_-=0$) or as the $\Delta\to 0^-$ limit of the sliced tachyonic orbits (one removes the circle $L_0=0$) and extend to them the previous parametrization and results.
\end{itemize}
We denote a generic orbit as $\mathcal O^{\spl{2}}_{\Delta,\sigma}$ where, for $\Delta>0$, the additional label $\sigma\in\{-1,+1\}$ distinguishes the two sheets, while, for $\Delta=0$, $\sigma\in\{-1,0,+1\}$ distinguishes the two cones and the origin. The interested reader can explore in greater depth the $\spl{2}$ group and representation theory by consulting the extensive literature on the subject (see~\cite{lang_discrete_1985, woodhouse_geometric_1992, rubilar_adjoint_2020, casselman_representations_nodate} for a selection of relevant references).

\subsection{Geometric quantization of coadjoint orbits}
\label{Subsec.geometricquantization}

Because they are symplectic, coadjoint orbits are good candidates to represent physically interesting phase spaces. Moreover, as they carry a transitive, Hamiltonian action of a symmetry group (the original group $G$) they provide irreducible representations of that group. In some sense, orbits provide an abstract class of symmetric phase spaces. Thus, results that we report in this section apply to coadjoint orbits as well as to general symplectic manifolds.
Similarly, at the quantum level, one expects the quantization of coadjoint orbit to provide Hilbert space representations of the symmetry group. This idea is at the core of Kirillov's program to obtain unitary group representations from the quantization of coadjoint orbits~\cite{kirillov_geometric_1990}.
Because of the lack of general results, the orbit method is not a proved theorem yet (except for some cases~\cite{kirillov_unitary_1962}), but it is still a powerful tool in representation theory~\cite{KirillovLectures, AuslanderKonstant, duflo_sur_1977, knapp, lahlali_coset_2024}.
We review here the basics of \textit{geometric quantization}, with the purpose of applying them in the following sections.

The starting point to geometrically quantize a coadjoint orbit $\mathcal O$ -- in this section, we omit writing the subscript -- is to introduce a complex line bundle $\mathcal L$ over it. Sections of this bundle serve as wavefunctions representing quantum states.\footnote{If $\pi:\mathcal L\to \mathcal O$ is the projection to the base manifold, a section $\psi$ is a map $\mathcal O\to\mathcal L$ such that $\pi\circ\psi=\mathrm{Id}_{\mathcal O}$} The two missing ingredients to complete \textit{prequantization} are a connection and a Hermitian structure on $\mathcal L$ that are compatible.
A hermitian structure is a smooth map $(\sbullet,\sbullet)$ that, for each point $\el{m}\in\mathcal O$, maps $\Gamma(\mathcal O,\mathcal L)\times \Gamma(\mathcal O,\mathcal L)\to\CC$ linearly in the second argument and antilinearly in the first. A connection is a rule that allows us to compare sections at different fibers in order to take derivatives in the space of sections $\Gamma(\mathcal O,\mathcal L)$. Specifically, given any (complex) vector field $\xi\in\mathfrak X_\CC(\mathcal O)$, the connection is a linear map $\Gamma(\mathcal O,\mathcal L)\to\Gamma(\mathcal O,\mathcal L)$ that satisfies the Leibniz rule
\begin{equation}
   \nabla_\xi (f\Psi)=\xi(f)\Psi+f\nabla_\xi \Psi, \quad\forall \Psi\in\Gamma(\mathcal O,\mathcal L),\,f\in C^\infty(\mathcal O,\CC).
\end{equation}
Compatibility between these two structures entails that
\begin{equation}
    \xi[(\Psi_1,\Psi_2)]=(\nabla_\xi \Psi_1,\Psi_2)+(\Psi_1,\nabla_\xi\Psi_2).
\end{equation}
If $\xi$ is a vector field that makes the l.h.s. vanish, this condition tells us that $i\nabla_\xi$ is an hermitian operator.

The basic request we make on a quantization prescription is that the Hilbert space must support an operator algebra that is associated with the Poisson algebra of classical observables in the sense that $[\hat{\mathcal X},\hat{\mathcal Y}]=i \widehat{\{\mathcal X,\mathcal Y\}}$. Because the Hamiltonian vector fields reproduce the Poisson algebra~\eqref{Eq.vfrepresentation}, the following association
\begin{equation}
    \mathcal X\to \hat{\mathcal X}=\mathcal X-i\nabla_{\xi_\mathcal X}
\end{equation}
is a suitable prescription as long as the curvature form of the connection is proportional to the KKS $R_\nabla=i\Omega$.
\begin{proof}
Direct evaluation shows that
\begin{equation}
    \begin{split}
        \frac{1}{i}[\hat{\mathcal X},\hat{\mathcal Y}]&=\frac{1}{i}[\mathcal X-i\nabla_{\xi_\mathcal X},\mathcal Y-i\nabla_{\xi_\mathcal Y}]=- \xi_\mathcal X(\mathcal Y)+ \xi_\mathcal Y(\mathcal X)+i[\nabla_{\xi_\mathcal X},\nabla_{\xi_\mathcal Y}]=\\
        &=2\{\mathcal X,\mathcal Y\}+i \nabla_{\Lieb{\xi_\mathcal X,\xi_\mathcal Y}}+i R_\nabla(\xi_\mathcal X,\xi_\mathcal Y)=2\{\mathcal X,\mathcal Y\}-i \nabla_{\xi_{\{\mathcal X,\mathcal Y\}}}+i R_\nabla(\xi_\mathcal X,\xi_\mathcal Y)=\\
        &=\widehat{\{\mathcal X,\mathcal Y\}}+(\{\mathcal X,\mathcal Y\}+i R_\nabla(\xi_\mathcal X,\xi_\mathcal Y)).
    \end{split}
\end{equation}
The last term in round brackets vanishes if and only if the curvature of $\nabla$ coincides with ($i$ times) the KKS Poisson brackets~\eqref{Eq.KKPoissonBrackets}, which proves the claim.
\end{proof}
In order to be promoted to the curvature form of a complex line bundle, the KKS form must satisfy an integrality condition in terms if its (de Rham) cohomology class, that is $[\Omega]\in H^2(\mathcal O,2\pi\,\ZZ)$.

Once prequantization is completed, the space one is dealing with is still too big. Since the orbits are symplectic manifolds, a section over it will generically depend on both \textit{positions} and \textit{momenta}, in contrast with the uncertainty principle. To fix this problem, one splits the variables in half and chooses the \textit{polarization} of wavefunctions to be either in position or momentum space. More covariantly, where there is no global way to make the separation of positions and momenta, a polarization is an involutive distribution $\mathcal P\subset T^\CC(\mathcal O)$ that is also Lagrangian, meaning that $\Omega(\xi_1,\xi_2)=0$ $\forall \xi_1,\xi_2\in \mathcal P$, and has maximal dimension (half the orbit dimension).

The choice of polarization is deeply related to the choice of connection because, after we construct the operators with the quantization map~\eqref{Eq.quantizationmap}, we must retain only those operators that preserve the polarization. In other words, the symplectic potential for $\Omega$ that one uses to build such a connection must be compatible with the polarization. We will see some examples of this in the following.

\section{Orbit analysis for corner symmetry groups}
\label{Sec.Orbits}

Before we analyze the coadjoint orbits of the QCS group, we review their construction for the ECS, and, for the sake of completeness, we summarize the findings of~\cite{ciambelli_universal_2023}.
In this section, for compatibility with the literature, we switch from the vectorial basis $\{\el{L}_0,\el{L_\pm}\}$ we introduced in Section~\ref{sec.UCS} to a matricial basis given by
\begin{equation}
    \el{J}^a{}_b=i\eta^{ij}(\tau_j)^a{}_b \el{L}_i,\qquad a,b=1,2,\quad i,j=0,+,-
\end{equation}
where the $\tau$ matrices can be taken to be $\tau_0=-\sigma_3$ and $\tau_\pm=\mp (\sigma_1\mp i\sigma_2)$. In this basis, the $\Liespl{2}$ Casimir is $\frac{1}{2}\el{J}^2$. Consistently, we also write
\begin{equation}
    \el{P}_a=\frac{1+i}{\sqrt{2}}(\el{T}_-,\el{T}_+),\qquad a,b=1,2.
\end{equation}
In this new basis, the $\mathfrak{h}_S$ commutation relations read
\begin{equation}
\label{Eq.realbasis}
    [\el{J}^a{}_b,\el{J}^c{}_d]=\delta^c_b \el{J}^a{}_d-\delta^a_d \el{J}^c{}_b,\quad  [\el{J}^a{}_b,\el{P}_c]=-\delta^a_c \el{P}_b+\frac{1}{2}\delta^a_b \el{P}_c, \quad [\el{P}_a, \el{P}_b]=0
\end{equation}
while the cubic Casimir is
\begin{equation}
\label{Eq.CasimirHS}
    \el{C}_3^{\h_S}=\frac{1}{4}\epsilon^{bc}(2\el{J}^a{}_b \el{P}_c \el{P}_a+ \el{P}_a \el{J}^a{}_b  \el{P}_c+\el{P}_c \el{P}_a \el{J}^a{}_b).
\end{equation}
where $\epsilon^{ab}$ is the inverse of the $\spl{2}$ invariant tensor
\begin{equation}
    \epsilon_{ab}=\mqty(0&1\\-1&0).
\end{equation}

\subsection{Coadjoint orbits of the extended corner symmetry group}
\label{Sec.OrbitsECS}

Being $5$-dimensional, the dual algebra $\h_S^*$ cannot be a symplectic manifold itself, which means that the maximal dimension of the $H_S$ orbits is $4$. We will see that this dimensionality reduction is due to the existence of the cubic Casimir in the algebra.

Given the real basis~\eqref{Eq.realbasis} in $\h_S$, we write a generic vector as $\el{X}=(\elg{\Theta},\elg{Z})=\theta^a{}_b\el{J}^b{}_a+\zeta^a\el{P_a}$.
For any $(h,n)\in H_S$, the adjoint action on $X$ is 
\begin{equation}
\label{Eq.adactionECS}
    \Ad{(h,n)}\el{X}=(h \theta h^{-1})^a{}_b\el{J}^b{}_a+(h^a{}_b\zeta^b-(h \theta h^{-1})^a{}_b n^b) \el{P}_a,
\end{equation}
where we wrote $h\in\spl{2}$ and $n\in\RR^2$ in their defining representations so that the action in the semidirect product $\spl{2}\ltimes\RR^2$ reduces to the natural action of matrices on vectors.

According to our convention, we write an element of the dual space as $\el{m}=(\el{j},\el{p})= J^a{}_b \elg{\theta}^b{}_a+P_a \elg{\zeta}^a$. We can deduce from~\eqref{Eq.adactionECS} that a group element acts on it as
\begin{equation}
\label{Eq.coadactionECS}
    \cAd{(h,n)}\el{m}=-(h J h^{-1})^a{}_b\elg{\theta}^b{}_a-P_a (h^{-1})^a{}_b (\elg{\theta}^b{}_c n^c-\elg{\zeta}^b).
\end{equation}
The push-forward of this action gives us the following transformations under the action of $\xi_{\el{X}}$
\begin{equation}
\label{Eq.OrbitsECS}
\begin{cases}
    \bbd_{\el{X}} J^a{}_b=\lbr J,\theta \rbr^a{}_b-P_b \zeta^a+\frac{1}{2} P_c \zeta^c \delta^a_b,\\
    \bbd_{\el{X}} P_a=P_b\,\theta^b{}_a. 
\end{cases}    
\end{equation}

Once we fix a basis in the algebra, we can use $\el{C}_3^{\h_S}$ to define a function on $\g^*$ that maps $\el{m}\to \mathcal C^{\h_S}_3(\el{m}):=\epsilon^{bc}J^a{}_b P_c P_a$ (notice that, unlike in~\eqref{Eq.CasimirHS}, $J^a{}_b$ and $P_c$ here are components in a given basis, hence they commute). The variation of this function is
\begin{equation}
\label{Eq.Casimirconservation}
   \bbd_{\el{X}}\mathcal C^{\h_S}_3(\el{m})= \epsilon^{bc}\pqty{\bbd_{\el{X}} J^a{}_b P_c P_a+J^a{}_b \bbd_{\el{X}} P_c P_a+J^a{}_b P_c \bbd_{\el{X}} P_a}=i(J^a{}_b P_d-J^a{}_d P_b)\theta^d{}_c \epsilon^{cb}=0,
\end{equation}
where the last equality follows because $\theta^d{}_c \epsilon^{cb}$ is symmetric.\footnote{
Any element $\el{B}$ of the universal enveloping algebra $U(\g)$ corresponds to a function in $\g^*$ via the following procedure. If we expand $\el{B}=\sum_n b_n \el{X}_{i_1}\otimes\dots\otimes \el{X}_{i_n}$, we construct the function $\mathcal B(\el{m})=\sum_n c_n m_{i_1}\dots m_{i_n}$ with $\el{X}_i(\el{m})=m_i$. The variation along the orbit can be computed with the Leibniz rule and considering that $\delta_{\el{X}} m_i=\pair{\cad{\el{X}} \el{m}}{\el{X}_i}=-\pair{\el{m}}{\ad{\el{X}}\el{X}_i}$
\begin{equation*}
\begin{split}
    \bbd_{\el{X}} \mathcal B(\el{m})&=-\sum_n b_n (\pair{\el{m}}{\ad{\el{X}} \el{X}_{i_1}}m_{i_2} \dots m_{i_n}+m_{i_1} \pair{\el{m}}{\ad{\el{X}} \el{X}_{i_2}}\dots m_{i_n}+\dots+m_{i_1}m_{i_2}\dots\pair{\el{m}}{\ad{\el{X}} \el{X}_{i_n}})=\\
    &=-\sum_n \pair{\el{m}\otimes \el{m}\otimes\dots\otimes \el{m}}{b_n \ad{\el{X}}(\el{X}_{i_1}\otimes \el{X}_{i_2}\otimes\dots\otimes \el{X}_{i_n})}=-\pair{\el{m}}{\ad{\el{X}} \el{B}}.
\end{split}
\end{equation*}
Then, if $\el{B}$ is a Casimir, it follows that $\mathcal B$ is invariant along the orbit.
} Orbits must therefore lie within the level sets of $\mathcal C_3^{H_S}$.
The KKS form on these orbits reads
\begin{equation}
    \Omega^{ECS}(\xi_{\el{X}},\xi_{\el{X}'})=-J^a{}_b \lbr \theta,\theta'\rbr^b{}_a-P_a(\theta^a{}_b\zeta'^b-\theta'^a{}_b\zeta^b)
\end{equation}
Ideally, one would like to invert the system~\eqref{Eq.OrbitsECS} and express the Lie algebra vector components in terms of the variations induced in the dual space element. Because $\mathcal C_3^{H_S}$ is constant along the orbit, the system has rank 4 and it is not invertible, but one can use~\eqref{Eq.Casimirconservation} to eliminate one variation and solve for the rest.

Let us consider again~\eqref{Eq.coadactionECS} and notice that the result only depends on $\el{p}$ through the combination $P_a(h^{-1})^a{}_b$, the ECS orbits -- just as the orbits of any semidirect group with a vectorial normal subgroup~\cite{oblak_bms_2017} -- form a fiber bundle over the $\spl{2}$ orbits in $\RR^{2*}$
\begin{equation}
    \mathcal{W}^{\spl{2}}_{\el{p}}=\{P_b h^b{}_a \elg{\zeta}^{a}| h\in\spl{2}\}
\end{equation}
Each fiber is a direct product of the cotangent space to this orbit with a coadjoint orbit
of the stabilizer $\spl{2}_{\el{p}}$ -- specifically, the one acting on the restriction of $\el{j}$ to the stabilizer itself. The orbits $\mathcal{W}^{\spl{2}}_{\el{p}}$ are very simple because, either $\el{p}=0$ and the orbit is trivial, or $\el{p}\ne 0$ and the orbit is the entire plane.
\begin{itemize}
    \item In the first case, the coadjoint action of the ECS reduces to $\cAd{(h,n)}\el{p}=(h^{-1}Jh)^a{}_b\elg{\theta}^b{}_a=\cAd{h}\el{j}$, and the stabilizer is trivially the entire $\spl{2}$. The coadjoint orbits of the ECS reduce to the $\spl{2}$ coadjoint orbits $\mathcal O^{\spl{2}}_{\el{j}}$, which are either $2$ or $0$ dimensional;
    \item In the second case, we can use a special linear element to set $\el{p}=(1,0)\in\RR^{2*}$. The stabilizer group is one dimensional, generated by the elements $\mqty(0&0\\1&0)\in\Liespl{2}_{(1,0)}$. The restriction of $\el{j}$ to the stabilizer is given by the equivalence class of $\Liespl{2}^*$ matrices with the same upper right entry $\bqty{\mqty(0&\delta\\0&0)}$. One can easily check that this class is preserved by the stabilizer coadjoint action. The total orbit dimension is the product of the dimensions of the base manifold and of its cotangent space $2\times 2=4$.
\end{itemize}

\subsection{Coadjoint orbits of the quantum corner symmetry group}
\label{Sec.OrbitsQCS}

After reviewing the coadjoint orbit analysis of the ECS, we can turn our attention to the quantum case, which for us means considering the group $H_Q=\widetilde{\spl{2}}\ltimes \HH_3$. The first difference with the previous case is that the normal subgroup is not abelian, which generally makes the analysis more complicated. However, the specific case we deal with is surprisingly simple, as we will see.
Let us recall that, in the defining representation of the Heisenberg group~\eqref{Eq.defrepHeisenberg}, the action of a $\spl{2}$ element is~\footnote{In the orbit analysis, we will consider only $\spl{2}$ and not its universal cover because, at the level of their action on the Heisenberg group, they are the same. Differences will appear when we realize the group action on some Hilbert space carrying a representation of the Heisenberg group.}
\begin{equation*}
    \mqty(u&v\\w&z)\lga \mqty(1&a&c\\0&1&b\\0&0&1)=\mqty(1&ua+vb&c+\frac{1}{2}(ua+vb)(wa+zb)-\frac{1}{2}ab\\0&1&wa+zb\\0&0&1).
\end{equation*}
For convenience, we report here the fundamental $\h_\mathrm{Q}$ Lie brackets in the matricial basis
\begin{equation}
    [\el{J}^a{}_b,\el{J}^c{}_d]=\delta^c_b \el{J}^a{}_d-\delta^a_d \el{J}^c{}_b,\quad [\el{J}^a{}_b,\el{P}_c]=-\delta^a_c \el{P}_b+\frac{1}{2}\delta^a_b \el{P}_c,\quad [\el{P}_a,\el{P}_b]=\epsilon_{ab}\el{E}.
\end{equation}
We define the dual basis $\{\elg{\theta}^a{}_b,\elg{\zeta}^a,\el{\sigma}\}$ of $\h_\mathrm{Q}^*$ by the non-trivial pairings
\begin{equation}
    \pair{\elg{\theta}^a{}_b}{\el{j}^c{}_d}=\delta^a_d\delta^c_b-\frac{1}{2}\delta^a_b\delta^c_d,\quad \pair{\elg{\zeta}^a}{\el{p}_b}=\delta^a_b,\quad \pair{\el{\sigma}}{\el{E}}=1.
\end{equation}
Therefore, if we write a generic element of $\h_\mathrm{Q}$ as $\el{X}=\theta^a{}_b \el{J}^b{}_a+\zeta^a \el{P}_a+\sigma \el{E}$ and an element of $\h_\mathrm{Q}^*$ as $\el{m}=J^a{}_b \elg{\theta}^b{}_a+P_a\elg{\zeta}^a+E\el{\sigma}$, their pairing gives
\begin{equation}
    \label{Eq.ElementPairing}
    \pair{\el{m}}{\el{X}}=J^a{}_b \theta^b{}_a+P_a\zeta^a+E\sigma.
\end{equation}

The adjoint action of the group on its algebra is
\begin{equation}
\label{eq.finiteQCSadjoint}
    \Ad{(h,n)}\el{X}=(h \theta h^{-1})^a{}_b\el{J}^b{}_a+(h^a{}_b\zeta^b-(h \theta h^{-1})^a{}_b n^b) \el{P}_a+\pqty{\sigma+n^a\epsilon_{ab}h^b{}_c\zeta^c-\frac{1}{2}n^a\epsilon_{ab}(h \theta h^{-1})^b{}_c n^c} \el{E},
\end{equation}
where $n^a=(n^1,n^2)$ are the components in the $\RR^2$ sector of $\HH^3$, while the last component -- $n^3$ -- does not contribute to the adjoint action.
Dualizing~\eqref{eq.finiteQCSadjoint}, we find the coadjoint action at group level
\begin{equation}
\label{Eq.orbitsQCS}
    \cAd{(h,n)}\el{m}=-(h J h^{-1})^a{}_b\elg{\theta}^b{}_a-P_c (h^{-1})^c{}_b n^a \elg{\theta}^b{}_a-\frac{1}{2}E n^c\epsilon_{bc} n^a \elg{\theta}^b{}_a+(P_b (h^{-1})^b{}_a-E n^b\epsilon_{ba})\elg{\zeta}^a+E \el{\sigma},
\end{equation}
and, at algebra level,
\begin{equation}
    \cad{\el{X}}\el{m}=(\lbr J,\theta\rbr^a{}_b-P_b\zeta^a)\el\theta^b{}_a+(P_a\theta^a{}_b-E \zeta^a \epsilon_{ab}) \el\zeta^b.
\end{equation}

From the cubic Casimir $\el{C}_3^{\h_Q}$, we construct the following function
\begin{equation}
    \mathcal C_3^{\h_Q}(\el{m})=\mathcal C_3^{\h_S}(\el{m})+2E\mathcal C_2^{\Liespl{2}}(\el{m})=\epsilon^{bc}J^a{}_b P_c P_a+E J^a{}_b J^b{}_a.
\end{equation}
which will be constant along the orbits. The presence of two Casimir functions (the second function is trivially $E$) implies that the $H_Q$ orbits are, at most, 4 dimensional.
We now wish to characterize all of them.

The first step would be to take the KKS form on the QCS coadjoint orbits
\begin{equation}
    \Omega^{\mathrm{QCS}}(\xi_{\el{X}},\xi_{\el{X}'})=-J^a{}_b \lbr\theta,\theta'\rbr^b{}_a-P_a(\theta^a{}_b \zeta'^b-\theta'^a{}_b \zeta^b)-E \zeta^a \epsilon_{ab} \zeta'^b,
\end{equation}
and express it in terms of coordinate differentials on the orbits by inverting the transformation induced by $\xi_{\el X}$ (and similar for $\xi_{\el X'}$)
\begin{equation}
\label{Eq.CoadjointFlow}
    \begin{cases}
       \bbd_{\el{X}} J^a{}_b=[J,\theta]^a{}_b-P_b\zeta^a+\frac{1}{2}\delta^a_b P_c\zeta^c,\\
       \bbd_{\el{X}} P_a=P_b\theta^b{}_a-E\zeta^b\epsilon_{ba},\\
       \bbd_{\el{X}} E=0.
    \end{cases}
\end{equation}
The computation appears to be laborious, but, inspired by the existence of the $\elg{\DL}$ generators~\eqref{Eq.Mgenerators}, we introduce the following polynomial functions
\begin{equation}
\label{Eq.Fcoordinates}
    \tilde J^a{}_b=J^a{}_b+\frac{1}{2E}\epsilon^{ac}P_c P_b.
\end{equation}
and use them instead of $J^a{}_b$ as local coordinates in any open set with $E\ne 0$. It is easy to check the following results:
\begin{equation}
    \delta_{\el{X}} \tilde J^a{}_b=\lbr\tilde J,\theta\rbr^a{}_b,
\end{equation}
\begin{equation}
    \tilde J^2({\el{m}})\equiv \tilde J^a{}_b({\el{m}}) \tilde J^b{}_a({\el{m}})=\frac{1}{E}\mathcal C^{\h_Q}_3({\el{m}}),
\end{equation}
which can be expressed as saying that $H_Q$ acts on the $\tilde J^a{}_b$ function only with its $\spl{2}$ part, preserving the associated Casimir function $\tilde J^2$. Therefore, whenever $E\ne 0$, the $H_Q$ orbits on $\tilde J^a{}_b$ are diffeomorphic to $\spl{2}$ orbits.

We can back up these observations by expressing the KKS form in the $\{\tilde J^a{}_b,P_a, E\}$ coordinates. What we see is that the symplectic form factorizes into two KKS forms, one for $\spl{2}$ and one for $\HH_3$
\begin{equation}
\label{Eq.FactorizationKKS}
    \Omega^{\mathrm{QCS}}_{\el{m}}(\xi_{\el X}, \xi_{\el X'})=-\underset{\spl{2}}{\underbrace{\tilde J^a{}_b \lbr\theta,\theta'\rbr^b{}_a}}+\underset{\HH_3}{\underbrace{\frac{1}{E}\bbd_{\el X} P_a \epsilon^{ab} \bbd_{\el X'} P_b}},
\end{equation}
implying that the QCS orbits factorize as well into a $\spl{2}$ coadjoint orbit on its algebra dual and a $\h_3$ coadjoint orbit, which is just a cotangent bundle. Therefore, we label a generic QCS orbits with $E\ne 0$ as
\begin{equation}
\mathcal{O}_{\Delta,\sigma,E}^{\mathrm{QCS}} =  \mathcal O^{\spl{2}}_{\Delta,\sigma}\times  \mathcal{O}^{\HH_3}_{E}.    
\end{equation}

The existence of functions $\tilde J^a{}_b$ that factorize the symplectic form can be seen as an application of Darboux's theorem. What is non-trivial is that these functions are well-defined in a very large region, which coincide with the region of interest where the QCS departs from the \textit{classical} regime described by the ECS.
Because of this decoupling, the QCS orbits for $E\ne 0$ are never 0-dimensional -- compatibly with the fact that setting $P_1=P_2=0$ is inconsistent with a non-zero commutator between them -- but they are 2- or 4- dimensional depending on whether $\tilde J^a{}_b$ vanishes as a matrix or not.

\section{Link with representations}
\label{Sec.Reps}

As we discovered in the last section, the coadjoint orbits of the QCS group factorize into independent orbits of the special linear group and the Heisenberg group when expressed in suitable coordinates. This structure greatly simplifies the quantization of QCS orbits, reducing it to the following three steps. First, one quantizes the orbits of the Heisenberg group described in Section~\ref{Sec.Method}, thereby obtaining the Hilbert space operators associated with the momentum map for the coordinates $P_a$. Second, one quantizes the orbits of $\spl{2}$ to obtain the corresponding operators for the coordinates $\tilde J^a{}_b$. Finally, since the orbits factorize in these coordinates, one constructs the Hilbert space associated with the full QCS orbit as the tensor product of the two Hilbert spaces obtained in the preceding steps. The operators for the original $\spl{2}$ coordinates can then be obtained by using the inverse of equation \eqref{Eq.Mgenerators}.

\subsection{Quantization of the Heisenberg orbits}

Let us apply the geometric quantization procedure we reviewed in Subsection~\ref{Subsec.geometricquantization} to the orbits we described in~\ref{Par.HeisenbergOrbits}. We consider the trivial line bundle over the coadjoint orbit with central element value $E$. Sections of this bundle are simply complex valued functions on the plane
\begin{equation}
    \phi: \mathcal{O}_{E}^{\HH_3}\cong \mathbb{R}^2 \longrightarrow \mathbb{C}.
\end{equation}
For readability and to match with the usual Schr\"odinger representation, we relabel the coordinates $(A,B,E)$ as $(x,p,e)$.
Since the bundle is trivial, we define the trivial point-wise hermitian structure
\begin{equation}\label{heisenberghermitian}
    \braket{\phi_1(x,p)}{\phi_2(x,p)}\defeq \overline{\phi_1(x,p)}\phi_2(x,p).
\end{equation}

The quantization map for an observable on the orbit $\mathcal X\in C^\infty(\RR^2)$ is given by
\begin{equation}\label{Eq.quantizationmap}
    \mathcal{X}\mapsto \hat{\mathcal{X}} = \mathcal{X} -i  \xi_{\mathcal{X}} - I_{\xi_{\mathcal{X}}}\theta^{\HH_3},
\end{equation}
where we choose $\theta^{\HH_3}=\frac{1}{e}p\,\bbd x$ as the connection. In particular, we will be interested in the quantization of the momentum map, since we know that these will furnish the irreducible representations. The Hamiltonian vector fields associated to them were calculated earlier
\begin{equation}
    \xi_{\el{A}} = -e \partial_{p},\quad \xi_{\el{B}} = e \partial_{x}.
\end{equation}
It is then straightforward to use equation \eqref{Eq.quantizationmap} to obtain the quantization of the momentum maps
\begin{equation}
\label{Eq.H3quantization}
    \hat{\mommap}_{\el{A}} = i e \,\partial_{p} + x,\quad
    \hat{\mommap}_{\el{B}} = -i e \,\partial_{x},\quad
    \hat{\mommap}_{\el E} = e.
\end{equation}
We now need to choose a polarization that is preserved by the observables defined above. It is easy to see that for the simple choice $\partial_{p} \phi = 0$, we get
\begin{equation}
    \partial_{p} \qty(\hat{\mommap}_{\el{A}}\phi) = \partial_{p} \qty(\hat{\mommap}_{\el{B}}
    \phi) = 0,
\end{equation}
consistently with the fact that $\theta^{\HH_3}(\partial_p)=0$.
Polarized wavefunctions are therefore complex valued function of $x\in\RR$, on which the observables act as 
\begin{equation}
\label{Eq.momentumoperatorheisenberg}
     \hat{\mommap}_{\el{A}} \phi(x) = x \phi(x),\quad \hat{\mommap}_{\el{B}}\phi(x) = -i e \partial_{x} \phi(x),\quad \hat{\mommap}_{\el E} \phi(x)= e \phi(x),
\end{equation}
which is the well known action of the Heisenberg group in quantum mechanics.

To construct the Hilbert space, we need to introduce a scalar product. Since we chose a real polarization, we need to use the \textit{half-form} construction \cite{bates_weinstein_quantization, Blau, carosso2018geometricquantization}. On a given orbit, we denote by $Q\subset \mathcal{O}_{E}^{\HH_3}$ the submanifold corresponding to the chosen polarization $\mathcal P = \mathrm{Span}\qty(\partial_{p})$ (an involutive distribution is integrable by Frobenius' theorem, $Q$ is the space of polarization leaves) and construct its complexified cotangent bundle $\mathrm{T}^*Q^{\mathbb{C}}$. Sections of this bundle are complexified forms on $\mathcal O^{\HH_3}_E$ that are by construction orthogonal to the polarization, that is $I_\xi \varpi=0$, $\forall\xi\in\mathcal P$.
More concretely, in $(x,p)$ coordinates, the choice of $\mathcal P=\text{Span}(\partial_p)$ leads to $\mathrm{T}^*Q^{\mathbb{C}} = \qty{\lambda\, \bbd x\mid \lambda \in C^{\infty}\qty(Q,\mathbb{C})}$.
We then introduce the square-root bundle $\delta_{\mathcal P} \longmapsto T^*Q$, whose sections are "half-forms" in the sense that $\delta_{\mathcal P}^{\otimes 2}\cong T^*Q^{\mathbb{C}}$.
The wave functions are then defined as products of the form $\tilde{\phi} = \phi \nu$, where $\phi$ is a complex line bundle section and $\nu \in \Gamma(\delta_{\mathcal P})$.
In the simple case where $Q$ is one-dimensional, a polarized~\footnote{The covariant derivative on products $\tilde\phi=\phi\nu$ is given by the Leibniz rule. Given the covariant derivative on $T^*Q^\CC$, $\nabla_\xi\varpi=I_\xi\dd\varpi$, one defines the covariant derivative on $\delta_\mathcal P$ such that
\begin{equation*}
    2\nu\otimes\nabla_\xi\nu=\nabla_\xi(\nu\otimes\nu)=I_\xi \dd(\nu\otimes\nu).
\end{equation*}
} wave function $\tilde\phi$ can be written as $\phi_{1/2}(x)\sqrt{\dd x}$, with a $\phi_{1/2}$ that transforms as a density of weight $1/2$:
\begin{equation}
    \phi_{1/2}(x)\to\pqty{\dv{x'}{x}}^{1/2}\phi_{1/2}(x').
\end{equation}

The hermitian structure \eqref{heisenberghermitian} for polarized wave functions now becomes
\begin{equation}
    \braket{\tilde{\phi}_1(x)}{\tilde{\phi}_2(x)} = \overline{\phi_1(x)} \phi_2(x) \bar{\nu}_1 \nu_2,
\end{equation}
and is a volume form on the polarized submanifold $Q$. It can therefore naturally be integrated to obtain the scalar product. Modulo possible irrelevant functions that can be reabsorbed in the definition of the wave functions, we obtain the standard $L^2\qty(\RR)$ scalar product
\begin{equation}
    \braket{\tilde{\phi}_1}{\tilde{\phi}_2} = \int_\RR \dd x\, \overline{\phi_1(x)}\phi_2(x), 
\end{equation}
for which it is easy to check that the quantized momentum map are Hermitian operators.

Finally, we define operators associated with $\el{T_-}$ and $\el{T_+}$ following Eq.~\eqref{eq.HeisenbergRotation}
\begin{equation}
    \hat\mommap_{\el{T}_\pm}=\frac{1}{\sqrt{2}}\pqty{\hat\mommap_{\el A}\pm i\hat\mommap_{\el{B}}}.
\end{equation}
Assuming $e>0$, we can use these to build creation and annihilation operators by~\footnote{For $e<0$, one just switches the roles of $\hat\mommap_{\el{T}_-}$ and $\hat\mommap_{\el{T}_+}$.}
\begin{equation}
\label{Eq.aidentification}
    \hat\mommap_{\el{T}_-}=\sqrt{e}a,\qquad \hat\mommap_{\el{T}_+}=\sqrt{e}a^\dagger,
\end{equation}
and check that $[a,a^\dagger]=\frac{1}{e}\hat\mommap_{\el{E}}=1$ indeed. We now introduce the Fock basis $\{\ket{k}\}$, constructed from the $\hat\mommap_{\el{A}}$ (position) eigenstate basis using Hermite polynomials~\footnote{The $k$-th Hermite polynomial is
\begin{equation*}
    H_k(x) = (-1)^k e^{x^2} \frac{\dd^k}{\dd x^k} e^{-x^2}.
\end{equation*}}
\begin{equation}
\label{Eq.h3fockbasis}
    \ket{k} = \frac{1}{\pi^{1/4} \sqrt{2^k k!}}\int_{-\infty}^\infty  \dd x \, H_k(x E) \, e^{-\frac{x^2 E}{2}} \ket{x}, 
\end{equation}
where we denoted $\braket{x}{\phi} \equiv \phi(x)$. On these states, we obtain the expected action of $a$ and $a^\dagger$:
\begin{align}
    a\ket{k}=\sqrt{k}\ket{k-1},\qquad
    a^\dagger\ket{n}=\sqrt{k+1}\ket{k+1}.
\end{align}

\subsection{Quantization of $\spl{2}$ orbits}
\label{Subsec.SL2Rquantization}

In this section, we present in detail the quantization of $\spl{2}$ massive orbits and make some comments about the other type of orbits.

\paragraph{Massive orbits}
Let us consider first the massive orbits $\mathcal O^{\spl{2}}_{\Delta,\pm}$ described in~\ref{Par.SLorbits}. As mentioned there, these are two-sheeted hyperboloids in $\Liespl{2}^*$ given by the relation $L^2=-\Delta$ with $\Delta>0$. For convenience, we write $\Delta=q^2$ with  $q>0$. 
In this example, we will describe the quantization of the upper sheet only, as the quantization of the lower sheet will be similar.
Every such orbit is diffeomorphic to the upper half complex plane $\HH^+=\{\tau\in\CC|\,\mathfrak I(\tau)>0\}$, with an explicit diffeomorphism given by
\begin{equation}
\label{Eq.OrbittoUHCdiffeo}
    \Phi:\HH^+\to\mathcal O^{\spl{2}}_{\Delta,+},\qquad (\tau,\bar\tau)\mapsto (L_+, L_-)=\pqty{q\frac{(\tau-i)(\bar\tau-i)}{2\,\mathfrak I(\tau)},q\frac{(\tau+i)(\bar\tau+i)}{2\,\mathfrak I(\tau)}}.
\end{equation}
We can explicitly compute and check that
\begin{equation}
    L_0=\pair{\Phi(\tau,\bar\tau)}{\el{L}_0}= q\frac{1+\abs{\tau}^2}{2\,\mathfrak I(\tau)}>0.
\end{equation}
The advantage of this construction is that the action of $\spl{2}$ on $\HH^+$ is naturally realized in terms of \textit{real} M\"obius transformations
\begin{equation}
    \tau\to \frac{a\tau +b}{c\tau+d}, \quad ad-bc=1,\quad a,b,c,d\in \RR.
\end{equation}
As these transformations do not mix $\tau$ and $\bar\tau$, there exists a polarization in $T^\CC\HH^+$ that is compatible with the $\spl{2}$ action: namely, the holomorphic polarization $\mathcal P = \mathrm{Span}\qty(\partial_{\bar\tau})$. The anti-holomorphic one $\bar{\mathcal P}$ would be preserved as well if we were to choose that. Notice that the holomorphic polarization does not have a counterpart on the hyperboloid tangent bundle, but only on its complexification. While this choice might appear strange -- even unmotivated in the framework that we have been describing -- it is necessary because there is no real polarization that is preserved by $\spl{2}$ on these orbits~\cite{woodhouse_geometric_1992}.

The pull-back of the KKS symplectic form to $\HH^+$ gives the unique (up to real constants) $\spl{2}$ invariant symplectic potential in $\Omega^2(\HH^+,\RR)$
\begin{equation}
\label{Eq.UHCsymplectic}
    \omega=\Phi^*\Omega^{\spl{2}}_{\Delta,+}=q\frac{\bbd\tau\wedge \bbd \bar\tau}{2i\,\mathfrak I(\tau)^2},
\end{equation}
which can be written as the double derivative of a K\"ahler potential
\begin{equation}
    \omega=-i\pdv[2]{\mathcal K}{\bar\tau}{\tau}\bbd\tau\wedge \bbd \bar\tau, \quad \mathcal K=-2 q\ln(\mathfrak I(\tau)).
\end{equation}
The existence of a K\"ahler structure naturally fits with the decomposition of $T^\CC \HH^+=\mathcal P\oplus \bar{\mathcal P}$. As we will work in the holomorphic polarization, we take the symplectic potential ($\omega=-\bbd\theta$) to be
\begin{equation}
    \theta=\frac{ q\bbd\tau}{\mathfrak I(\tau)},
\end{equation}
which trivially satisfies $\theta(\mathcal P)=0$.
Given the push-forward of the fundamental vector fields $\xi_{\el{L}_i}$ via the inverse map
\begin{subequations}
\begin{align}
    &\Phi^{-1}_*\xi_{\el{L}_0}=\frac{1}{2}\pqty{(1+\tau^2)\partial_\tau+(1+\bar\tau^2)\partial_{\bar\tau}},\\
    &\Phi^{-1}_*\xi_{\el{L}_-}=\frac{1}{2}\pqty{(\tau+i)^2\partial_\tau+(\bar\tau+i)^2\partial_{\bar\tau}}
    ,\\
    &\Phi^{-1}_*\xi_{\el{L}_+}=\frac{1}{2}\pqty{(\tau-i)^2\partial_\tau+(\bar\tau-i)^2\partial_{\bar\tau}}
    ,
\end{align}
\end{subequations}
we construct the quantum operators acting on holomorphic sections of the line bundle $\psi\in\Gamma_{\mathcal P}(\HH^+,\mathfrak L)$ via the geometric quantization prescription:
\begin{subequations}
\label{Eq.MassOrbitquantization}
    \begin{align}
        &\hat\mommap_{\el{L}_{0}}\psi=-i\frac{1+\tau^2}{2}\pdv{\psi}{\tau}-i q\tau\psi,\\
        &\hat\mommap_{\el{L}_{+}}\psi=-i \frac{(\tau-i)^2}{2}\pdv{\psi}{\tau}-i q(\tau-i)\psi,\\
        &\hat\mommap_{\el{L}_{-}}\psi=-i \frac{(\tau+i)^2}{2}\pdv{\psi}{\tau}-i q(\tau+i)\psi.
     \end{align}
\end{subequations}

To turn the space of holomorphic sections into a Hilbert space, we need to introduce a positive definite scalar product. On $\HH^+$, there exists a family of $\spl{2}$ covariant measures given by~\cite{lang_sl2r_1985}
\begin{equation}
    \dd\mu_m=\dd\mu\, e^{K}=\frac{\dd\tau\dd\bar\tau}{\mathfrak I(\tau)^2}\mathfrak I(\tau)^{m}, \quad m>1,
\end{equation}
where $K=m\ln(\mathfrak I(\tau))$ is a metric on the line bundle. If we want $-\theta$ to be the connection compatible with this metric, we must have $K=-\mathcal K$, which means $m=2q$.
The Hilbert space that we associate to an orbit with Casimir value $-q^2$ is $\mathrm{L}^2(\HH^+,\dd\mu_{2q})\subset \Gamma_{\mathcal P}(\HH^+,\mathfrak L)$, where the inner product is given by
\begin{equation}
    \braket{\psi_1}{\psi_2}=\int_{\HH^+}\dd\mu_{2q} \overline{\psi_1(\tau)}{\psi_2}(\tau).
\end{equation}
The action of $\spl{2}$ preserves this product provided that the wavefunctions transform as the component of differential forms of degree $q$, i.e.
\begin{equation}
\label{Eq.finitetrafo}
    \psi(\tau)\to \pqty{\frac{1}{c\tau+d}}^{2q}\psi\pqty{\frac{a\tau+b}{c\tau+d}}.
\end{equation}
With respect to this inner product, the operators in~\eqref{Eq.MassOrbitquantization} satisfy $\hat{\mathcal J_{\el{L}_0}}^\dagger=\hat{\mathcal J_{\el{L}_0}}$ and $\hat{\mathcal J_{\el{L}_-}}^\dagger=\hat{\mathcal J_{\el{L}_+}}$, which means that they provide a unitary representation of $\widetilde{\spl{2}}$. If the transformation~\eqref{Eq.finitetrafo} is single-valued (which happens for $q\in \ZZ/2$), the Hilbert space carries a representation of the non-extended $\spl{2}$. This construction holds for $q>1/2$, whereas there exists an alternative unitary structure for $0<q\le 1/2$~\cite{witten_coadjoint_1988}.

If we denote by $\ket{n}\equiv\ket{n,q}$ the state such that
\begin{equation}
\label{Eq.sl2rfockbasis}
    \braket{\tau}{n}=c_n\pqty{\frac{\tau-i}{\tau+i}}^{n}(1+\tau^2)^{-q},
\end{equation}
where $c_n$ is a suitable normalization constant ($\abs{c_n}<\infty$~\cite{lang_discrete_1985}), we can show that
\begin{subequations}
\label{Eq.momentumoperatorSL}
\begin{align}
    &\hat\mommap_{\el{L}_{0}}\ket{n}= n\ket{n},\\
    &\hat\mommap_{\el{L}_{-}}\ket{n}= \sqrt{n(n-1)+q(1-q)}\ket{n-1},\\
    &\hat\mommap_{\el{L}_{+}}\ket{n}= \sqrt{n(n+1)+q(1-q)}\ket{n+1}.
\end{align}
\end{subequations}
This prove that $\mathrm{L}^2(\HH^+,\dd\mu_{2q})$ with $q>1/2$ coincides with $\mathcal H_{q,\mu=q}$, the $\widetilde{\spl{2}}$ positive discrete series representation space. Unlike for $\su{2}$, the discreteness of these representations does not come from the integrality conditions, but from the single-valuedness of~\eqref{Eq.finitetrafo}.
Finally, one can check that the Casimir $\el{L}^2$ is represented by a multiple of the identity
\begin{equation}
\label{Eq.SL2CasimirRep}
    \pqty{-\hat\mommap_{\el{L}_{0}}^2+\frac{1}{2}(\hat\mommap_{\el{L}_{+}}\hat\mommap_{\el{L}_{-}}+\hat\mommap_{\el{L}_{-}}\hat\mommap_{\el{L}_{+}})}\ket{n}=q(1-q)\ket{n},
\end{equation}
finding that the Casimir value $-q^2$ is corrected to $q(1-q)$ at the quantum level.

\paragraph{Tachyonic orbits}
These orbits are obtained when $\Delta<0$. Topologically, they are cylinder and can thus be seen as the cotangent bundle of a circle $\mathcal O^{\spl{2}}=T^*\SSS^1$~\cite{plyushchay_quantization_1993}. Using the symplectic form given in~\eqref{Eq.tachyonicKKS}, we get the symplectic form on the cylinder
\begin{equation}
\label{Eq.CylinderForm}
    \omega=\bbd L_0\wedge \bbd\varphi.
\end{equation}
Cotangent bundles are easy to quantize: the Hilbert space is just the space of complex functions of the base manifold: the circle, in this case. As for the Heisenberg orbits quantization, we need to introduce half-forms here as well, which means that we actually consider objects $\psi(\varphi)\nu$, where $\psi\in L^2(\SSS,\dd\varphi)$ and $\nu\in\Gamma(\delta_{\mathcal P})$. Choosing the polarization $\theta=(\mu-L_0) \bbd \varphi$, we quantize the functions $\mommap_{\el{L}_i}$ and get
\begin{subequations}
\begin{align}
    &\hat \mommap_{\el{L}_0}\psi=i\pdv{\psi}{\varphi}+\mu\psi,\\
    &\hat \mommap_{\el{L}_\pm}\psi=ie^{\mp i\varphi}\pqty{i\pdv{\psi}{\varphi}+\mu\psi\pm i \pqty{s-\frac{i}{2}}\psi}.
\end{align}
\end{subequations}
These operators satisfy
\begin{equation}
    \pqty{-\hat\mommap_{\el{L}_{0}}^2+\frac{1}{2}(\hat\mommap_{\el{L}_{+}}\hat\mommap_{\el{L}_{-}}+\hat\mommap_{\el{L}_{-}}\hat\mommap_{\el{L}_{+}})}\psi=\pqty{s^2+\frac{1}{4}}\psi,
\end{equation}
which shows that the Casimir is shifted from $s^2$ to $s^2+1/4$ at the quantum level. This value of the Casimir corresponds to a representation in the continuous principal series. We can show this explicitly introducing the states $\ket{n}\equiv\ket{n,\frac{1}{2}+is}$ by
\begin{equation}
\label{Eq.sl2rfockbasis2}
    \braket{\phi}{n}=\frac{1}{\sqrt{2\pi}}e^{i (n-\mu)\phi}, \qquad\text{with}\quad n\in\mu+\ZZ.
\end{equation}
These are such that
\begin{subequations}
\label{Eq.momentumoperatorSL2}
\begin{align}
    &\hat\mommap_{\el{L}_{0}}\ket{n}= n\ket{n},\\
    &\hat\mommap_{\el{L}_{-}}\ket{n}= \sqrt{n(n-1)+s^2+1/4}\ket{n-1},\\
    &\hat\mommap_{\el{L}_{+}}\ket{n}= \sqrt{n(n+1)+s^2+1/4}\ket{n+1},
\end{align}
\end{subequations}
proving that their span is the Hilbert space $\mathcal H_{1/2+is,\mu}$.

\paragraph{Light-like orbits}
Unlike the others, the quantization of these orbits is elusive~\cite{witten_coadjoint_1988}. They lack a (evident, at least) K\"ahler or cotangent bundle structure. Moreover, they do not even have a scale that might work as a coupling constant, hindering perturbative quantization

\paragraph{Trivial orbit}
The orbit is a single point at $L_i=0$, corresponding to the trivial representation -- the only possible finite-dimensional unitary representation of a non-compact Lie group.
In the context of the QCS, where the functions that transform under $\spl{2}$ are the coordinates $\tilde J^a{}_b$, this orbit will correspond to the metaplectic representation, and the operators corresponding to $J^a{}_b$ will be constructed as composite operators from the quantum version of $T_+$, $T_-$ and $E$. 

\subsection{Quantization of the QCS orbits}

We are now ready to perform the last step of the construction. As mentioned, because of the orbit factorization, we expect the Hilbert space to be given by the tensor product of the Hilbert spaces we constructed in the last two sections. If $\mathcal H_{\Delta,\sigma}$ is the Hilbert space we constructed out of the orbit $\mathcal O^{\spl{2}}_{\Delta,\sigma}$ (for the $\tilde J^a{}_b$ coordinates) and $\Fock_e$ is the Hilbert (Fock) space we constructed out of the orbit $\mathcal O^{\HH_3}_E$, the total Hilbert space is simply the product
\begin{equation}\label{eq:qcshilbertspace}
    \mathcal H_{\Delta,\sigma,e}=\mathcal H_{\Delta,\sigma}\otimes \Fock_e.
\end{equation}
Indeed, geometric quantization can be performed independently on the two orbits and, after the vector space is constructed, the correct scalar product is just the one coming from the tensor product. In the following, we are going to assume $e\equiv E>0$ -- the case $e<0$ is not very different.

In the previous sections, we constructed operators that represent the elements of $\Liespl{2}$ and $\h_3$ on the two sides of the product~\eqref{eq:qcshilbertspace} (see Eq.s~\eqref{Eq.momentumoperatorheisenberg}, \eqref{Eq.momentumoperatorSL} and \eqref{Eq.momentumoperatorSL2}). The natural extension to the total Hilbert space is
\begin{equation}
\label{Eq.qcsdecoupledoperators}
\QCSop_{\elg{\DL}_i} = \hat{\mommap}_{\el{L}_i}\otimes  \mathds{1} ,
\quad 
\hat{\mommap}_{{\el{T}_\pm}}^{\mathrm{QCS}} =  \mathds{1} \otimes \hat{\mommap}_{\el{T}_\pm},
\quad
\hat{\mommap}_{\el{E}}^{\mathrm{QCS}} =  \mathds{1} \otimes \hat{\mommap}_{\el{E}},
\end{equation}
where it is important to notice that we associate the first set of operators to the $\el{\DL}_i$ elements in the universal enveloping algebra~\eqref{Eq.Mgenerators} rather than the standard generators. In fact, the possibility to write the operators in this tensor product form reflects the orbit factorization expressed in $\tilde J^a{}_b$ coordinates, which in turn follows from the commutativity of the modified generators $\elg{\DL}_i$ with $\el{T}_\pm$ (see Eq.~\eqref{Eq.AlegbraFactorization}).

In this picture, we can obtain the the quantum version of the original $\spl{2}$ coordinates (those that do not decouple from the Heisenberg sector) by combining the operators that we just constructed. 
Following equation \eqref{Eq.Mgenerators}, we define them as
\begin{subequations}
\begin{align}
   \QCSop_{\el{L}_0} &=\QCSop_{\elg{\DL}_0}+\frac{1}{4e}\pqty{\QCSop_{\el{T}_-}\QCSop_{\el{T_+}}+\QCSop_{\el{T}_+}\QCSop_{\el{T}_-}}= \hat{\mommap}_{\el{L}_0}\otimes \mathds{1} + \mathds{1}\otimes \frac12\qty(a^\dagger a + \frac12),\\ 
   \QCSop_{\el{L_-}} &= \QCSop_{\elg{\DL}_-}+\frac{1}{2e}\QCSop_{\el{T}_-}\QCSop_{\el{T_-}}= \hat{\mommap}_{\el{L}_-}\otimes \mathds{1} + \mathds{1}\otimes\frac12 a a,\\
   \QCSop_{\el{L_+}} &=\QCSop_{\elg{\DL}_+}+\frac{1}{2e}\QCSop_{\el{T}_+}\QCSop_{\el{T_+}}= \hat{\mommap}_{\el{L}_+}\otimes \mathds{1} + \mathds{1}\otimes\frac12 a^\dagger a^\dagger,
\end{align}
\end{subequations}
where we have used the identification \eqref{Eq.aidentification}.

The action of the QCS operators on the Hibert space \eqref{eq:qcshilbertspace} is naturally expressed in the basis $\ket{n,k}\equiv \ket{n}\otimes \ket{k}$, where $\ket{n}$ is the basis of $\mathcal H_{\Delta,\sigma}$ introduced in~\eqref{Eq.sl2rfockbasis} (for the massive orbits) or in~\eqref{Eq.sl2rfockbasis2} (for the tachyonic orbits), and $\ket{k}$ is the basis of $\Fock_e$ defined that we just defined:
\begin{subequations}
\begin{align}
    &\QCSop_{\el{T}_-}\ket{n,k}=\sqrt{e}\sqrt{k}\ket{n,k-1},\\
    &\QCSop_{\el{T}_+}\ket{n,k}=\sqrt{e}\sqrt{k+1}\ket{n,k+1},\\
    &\QCSop_{\el{L}_0}\ket{n,k}= \qty(n + \frac{2k + 1}{4})\ket{n,k},\\
    &\QCSop_{\el{L}_-}\ket{n,k}=-\sqrt{n(n-1)+q(1-q)}\ket{n-1,k}+\frac{1}{2}\sqrt{k(k-1)}\ket{n,k-2},\\
    &\QCSop_{\el{L}_+}\ket{n,k}=- \sqrt{n(n+1)+q(1-q)}\ket{n+1,k}+\frac{1}{2}\sqrt{(k+1)(k+2)}\ket{n,k+2},
\end{align}
\end{subequations}
which correspond, up to an automorphism in the $\QCSop_{\elg{\DL}_i}$ algebra, to the representations of~\cite{varrin_physical_2024} that we reported in~\eqref{Eq.VarrinReps}.
One can check that these operators satisfy the commutation relations given by the QCS algebra. Moreover, if we define the Casimir operator,~\footnote{The natural definition of the QCS Casimir operator comes from~\eqref{eq.SL2Casimir}, \eqref{eq.HSCasimir}, and \eqref{eq.HQCasimir}
\begin{equation*}
\begin{split}
    \QCSop_{\el{C}^{\h_Q}_3}:=&-\QCSop_{\el{L}_+}{\QCSop_{\el{T}_-}}{}^2-\QCSop_{\el{L}_-}{\QCSop_{\el{T}_+}}{}^2+\pqty{\QCSop_{\el{L}_0}+\frac{3}{8}i}\QCSop_{\el{T}_+}\QCSop_{\el{T}_-}+\pqty{\QCSop_{\el{L}_0}-\frac{3}{8}i}\QCSop_{\el{T}_-}\QCSop_{\el{T}_+}+\\
    &+2e\pqty{-{\QCSop_{\el{L}_0}}{}^2+\frac{1}{2}\QCSop_{\el{L}_+}\QCSop_{\el{L}_-}+\frac{1}{2}\QCSop_{\el{L}_-}\QCSop_{\el{L}_+}}.
    \end{split}
\end{equation*}} we can see that
\begin{equation}
    \QCSop_{\el{C}^{\h_Q}_3}\ket{n,k}=2eq(1-q) \ket{n,k}.
\end{equation}
gives the same result as
\begin{equation}
    2e\pqty{-\QCSop_{\elg{\DL}_0}{}^2+\frac{1}{2}\pqty{\QCSop_{\elg{\DL}_+}\QCSop_{\elg{\DL}_-}+\QCSop_{\elg{\DL}_-}\QCSop_{\elg{\DL}_+}}}\ket{n,k},
\end{equation}
as expected from~\eqref{Eq.QCScasimirM} and~\eqref{Eq.SL2CasimirRep}.

\section{Conclusion}
\label{sec.Conclusion}
The orbit method has long been an exceptional tool to study the representation theory of Lie groups. While the correspondence between coadjoint orbits and unitary irreducible representations is proven only for unipotent Lie groups, its application to the Virasoro and Kac-Moody algebras has proven very useful in the context of lower-dimensional quantum gravity \cite{witten_coadjoint_1988,alekseev_quantization_1988,Rai:1989js,Brensinger:2017gtb,Caputa:2018kdj,Erdmenger:2020sup}. Even if some of these works cited above do address semidirect products with non-abelian normal subgroups, this case remains less well understood than its abelian counterpart. The present work thus contributes to the growing understanding of the non-abelian case. 

In this work, we have discussed the emergence of corner symmetries in gravitational theories and argued for their significance within the bottom-up approach of the corner proposal for quantum gravity. We began reviewing the key ideas and mathematical framework underlying the orbit method. The application of this method to corner symmetries was initiated in \cite{ciambelli_isolated_2021} with the study of the orbits of the local, finite-dimensional part of the extended corner symmetry algebra. We reviewed those results in Section~\ref{Sec.OrbitsECS} and then tackled their generalization to the quantum case.

The main result of this paper is the demonstration that, starting from the standard Lie algebra basis of the semi-direct product group $\spl{2}\ltimes\HH_3$ (where $\HH_3$ is the Heisenberg group in 1d) and the associated coordinates on its dual algebra, one can construct mixed coordinate functions that both satisfy an $\Liespl{2}$ algebra under the Poisson bracket and decouple from the Heisenberg sector.
The existence of these coordinates shows that the orbits of the quantum corner symmetry (QCS) group decompose into independent $\spl{2}$ and Heisenberg orbits. This factorization greatly simplifies the quantization procedure, as each sector can be quantized independently provided that we express the $\spl{2}$ orbits in terms of the mixed coordinates.
Operators associated with the original $\spl{2}$ sector of the QCS group can then be recovered by simply undoing the mixing.

The resulting representations are in complete correspondence with those found in~\cite{varrin_physical_2024}, modulo the complementary series of $\mathfrak{sl}(2,\RR)$, which cannot be obtained through orbit quantization -- reflecting the fact that they do not contribute to the Plancherel measure \cite{harishchandra_harmonic_1977}. In that work, one of the authors derived the representations using Mackey's theory of induced representations, relying on the fact that the little group -- the subgroup that preserves the (equivalence class of) representations of the normal subgroup -- is the full $\widetilde{\spl{2}}$. In this case, the representations can be realized on the tensor product of the representation spaces of the Heisenberg group and $\widetilde{\spl{2}}$. The fact that we recovered the same results suggests that this property can be understood as the algebraic counterpart of the orbit factorization. The results of this work offers a promising new approach to studying the coadjoint orbits of semi-direct product groups with non-abelian normal subgroups that admit a global little group.

Understanding the structure of coadjoint orbits is particularly significant in the context of determining the Plancherel measure for the QCS group. This measure is believed to play a fundamental role when corners are interpreted as entangling surfaces of local subregions, with important implications for entanglement entropy \cite{Ciambelli2025}.
In the mathematical literature, the QCS group is known as the Jacobi group. While its representation theory has been extensively studied \cite{berndt_elements_1998}, a complete determination of the Plancherel decomposition of square-integrable functions -- with respect to the Haar measure -- has not yet been achieved. Our findings suggest that obtaining the Plancherel measure via the coadjoint orbit method may be significantly more straightforward than expected.

The successful application of the orbit method to derive the unitary irreducible representations of the quantum corner symmetry group in two dimensions is highly encouraging and opens the way for its generalization to higher dimensions. Upon reintroducing diffeomorphisms, the representation theory is expected to become significantly more intricate, 
potentially limiting the applicability of algebraic techniques. In this context, the orbit method may prove to be the only viable approach. It is therefore crucial that the method has been validated in the two-dimensional case.

Finally, the coadjoint orbits can be utilized to construct world-line actions associated with the quantum corner symmetry group~\cite{Alekseev:1988ce, Alekseev:1988vx, Aratyn:1990dj, Charles:1999qq}. This, in turn, allows us to interpret the representations of the QCS group as actual quantized particles. Understanding the nature of these \textit{QCS particles} and developing an associated scattering theory is an exciting direction for future research. Several interesting questions come to mind: Do these particles correspond to causal diamonds? What genuinely quantum observables can be defined in this framework? Can QCS particle scattering be associated with a topology change? In what sense these particles constitute fundamental geometrical degrees of freedom?
Is the entanglement between them related to spacetime connectivity?
Answering these and other questions constitutes an important step in advancing the corner proposal program.

\section*{Acknowledgements}

The authors are grateful to Luca Ciambelli and Jerzy Kowalski-Glikman for their valuable guidance from the start to the end of this project.
GN expresses its gratitude to Matteo Bruno, Simon Langenscheidt and Beniamino Valsesia for their support during the writing of this manuscript, and to Wolfgang Wieland and Muxin Han for insightful conversations about the big picture.
GN also thanks INFN for financial support and Perimeter Institute for hospitality.
LV is also grateful to Laurent Freidel and Robert Leigh for insightful discussions and thanks Perimeter Institute for its hospitality.
Research at Perimeter Institute is supported in part by the Government of Canada through the Department of Innovation, Science and Economic Development Canada and by the Province of Ontario through the Ministry of Colleges and Universities.

\bibliographystyle{JHEP}
\bibliography{main}

\providecommand{\href}[2]{#2}\begingroup\raggedright\begin{thebibliography}{10}

\bibitem{lagrange1811}
J.L.~Lagrange, \emph{Mecanique analytique}, vol.~1, Mme. Ve. Courcier, Paris
  (1811).

\bibitem{lagrange1815}
J.L.~Lagrange, \emph{Mecanique analytique}, vol.~2, Mme. Ve. Courcier, Paris
  (1815).

\bibitem{hamilton1834}
W.R.~Hamilton, \emph{Second essay on a general method in dynamics},
  {\emph{Philosophical Transactions of the Royal Society} (1834) 95}.

\bibitem{hamilton1835}
W.R.~Hamilton, \emph{On a general method in dynamics}, {\emph{Philosophical
  Transactions of the Royal Society} (1835) 247}.

\bibitem{jacobi1866}
C.G.J.~Jacobi, \emph{Vorlesungen {\"u}ber Dynamik}, Georg Reimer, Berlin
  (1866).

\bibitem{noether1918}
E.~Noether, \emph{Invariante {V}ariationsprobleme}, {\emph{Nachrichten von der
  Gesellschaft der Wissenschaften zu Göttingen, Mathematisch-Physikalische
  Klasse} (1918) 235}.

\bibitem{Wigner:1939cj}
E.P.~Wigner, \emph{{On Unitary Representations of the Inhomogeneous Lorentz
  Group}}, \href{https://doi.org/10.2307/1968551}{\emph{Annals Math.}
  {\bfseries 40} (1939) 149}.

\bibitem{Souriau1958}
J.-M.~Souriau, \emph{Mati\`ere parfaite en relativit\'e g\'en\'erale},
  {\emph{S\'eminaire Janet. M\'ecanique analytique et m\'ecanique c\'eleste}
  {\bfseries 3} (1959-1960) 1}.

\bibitem{kirillov_unitary_1962}
A.A.~Kirillov, \emph{Unitary representations of nilpotent lie groups},
  \href{https://doi.org/10.1070/RM1962v017n04ABEH004118}{\emph{Russian
  Mathematical Surveys} {\bfseries 17} (1962) 53}.

\bibitem{souriau1970structure}
J.~Souriau, \emph{Structure des syst{\`e}mes dynamiques: ma{\^\i}trises de
  math{\'e}matiques}, Collection Dunod Universit{\'e}, Dunod (1970).

\bibitem{Kostant1970}
B.~Kostant, \emph{Quantization and unitary representations},  in \emph{Lectures
  in Modern Analysis and Applications III}, C.T.~Taam, ed., (Berlin,
  Heidelberg), pp.~87--208, Springer Berlin Heidelberg, 1970.

\bibitem{kirillov_geometric_1990}
A.A.~Kirillov, \emph{Geometric {Quantization}},  in \emph{Dynamical {Systems}
  {IV}: {Symplectic} {Geometry} and its {Applications}}, V.I.~Arnol’d and
  S.P.~Novikov, eds., (Berlin, Heidelberg), pp.~137--172, Springer (1990),
  \href{https://doi.org/10.1007/978-3-662-06793-2_2}{DOI}.

\bibitem{KirillovLectures}
A.A.~Kirillov, \emph{Lectures on the Orbit Method}, vol.~64 of \emph{Graduate
  Studies in Mathematics}, American Mathematical Society, Providence, RI
  (2004).

\bibitem{regge_role_1974}
T.~Regge and C.~Teitelboim, \emph{Role of surface integrals in the
  {Hamiltonian} formulation of general relativity},
  \href{https://doi.org/10.1016/0003-4916(74)90404-7}{\emph{Annals of Physics}
  {\bfseries 88} (1974) 286}.

\bibitem{donnelly_local_2016}
W.~Donnelly and L.~Freidel, \emph{Local subsystems in gauge theory and
  gravity}, \href{https://doi.org/10.1007/JHEP09(2016)102}{\emph{Journal of
  High Energy Physics} {\bfseries 2016} (2016) 102}.

\bibitem{speranza_local_2018}
A.J.~Speranza, \emph{Local phase space and edge modes for
  diffeomorphism-invariant theories},
  \href{https://doi.org/10.1007/JHEP02(2018)021}{\emph{Journal of High Energy
  Physics} {\bfseries 2018} (2018) 21}.

\bibitem{freidel_edge_2020}
L.~Freidel, M.~Geiller and D.~Pranzetti, \emph{Edge modes of gravity. {Part}
  {I}. {Corner} potentials and charges},
  \href{https://doi.org/10.1007/JHEP11(2020)026}{\emph{Journal of High Energy
  Physics} {\bfseries 2020} (2020) 26}.

\bibitem{donnelly_gravitational_2021}
W.~Donnelly, L.~Freidel, S.F.~Moosavian and A.J.~Speranza, \emph{Gravitational
  edge modes, coadjoint orbits, and hydrodynamics},
  \href{https://doi.org/10.1007/JHEP09(2021)008}{\emph{Journal of High Energy
  Physics} {\bfseries 2021} (2021) 8}.

\bibitem{Speranza:2022lxr}
A.J.~Speranza, \emph{{Ambiguity resolution for integrable gravitational
  charges}}, \href{https://doi.org/10.1007/JHEP07(2022)029}{\emph{JHEP}
  {\bfseries 07} (2022) 029}
  [\href{https://arxiv.org/abs/2202.00133}{{\ttfamily 2202.00133}}].

\bibitem{ciambelli_isolated_2021}
L.~Ciambelli and R.G.~Leigh, \emph{Isolated surfaces and symmetries of
  gravity}, \href{https://doi.org/10.1103/PhysRevD.104.046005}{\emph{Physical
  Review D} {\bfseries 104} (2021) 046005}.

\bibitem{ciambelli_embeddings_2022}
L.~Ciambelli, R.G.~Leigh and P.-C.~Pai, \emph{Embeddings and {Integrable}
  {Charges} for {Extended} {Corner} {Symmetry}},
  \href{https://doi.org/10.1103/PhysRevLett.128.171302}{\emph{Physical Review
  Letters} {\bfseries 128} (2022) 171302}.

\bibitem{freidel_extended_2021}
L.~Freidel, R.~Oliveri, D.~Pranzetti and S.~Speziale, \emph{Extended corner
  symmetry, charge bracket and {Einstein}’s equations},
  \href{https://doi.org/10.1007/JHEP09(2021)083}{\emph{Journal of High Energy
  Physics} {\bfseries 2021} (2021) 83}.

\bibitem{geiller_extended_2020}
M.~Geiller and P.~Jai-akson, \emph{Extended actions, dynamics of edge modes,
  and entanglement entropy},
  \href{https://doi.org/10.1007/JHEP09(2020)134}{\emph{Journal of High Energy
  Physics} {\bfseries 2020} (2020) 134}.

\bibitem{Carrozza:2022xut}
S.~Carrozza, S.~Eccles and P.A.~Hoehn, \emph{{Edge modes as dynamical frames:
  charges from post-selection in generally covariant theories}},
  \href{https://doi.org/10.21468/SciPostPhys.17.2.048}{\emph{SciPost Phys.}
  {\bfseries 17} (2024) 048}
  [\href{https://arxiv.org/abs/2205.00913}{{\ttfamily 2205.00913}}].

\bibitem{Freidel:2023bnj}
L.~Freidel, M.~Geiller and W.~Wieland, \emph{{Corner Symmetry and Quantum
  Geometry}},  in \emph{{Handbook of Quantum Gravity}}, C.~Bambi, L.~Modesto
  and I.~Shapiro, eds., pp.~1--36 (2024)
  [\href{https://arxiv.org/abs/2302.12799}{{\ttfamily 2302.12799}}].

\bibitem{ciambelli_universal_2023}
L.~Ciambelli and R.~G.~Leigh, \emph{Universal corner symmetry and the orbit
  method for gravity},
  \href{https://doi.org/10.1016/j.nuclphysb.2022.116053}{\emph{Nuclear Physics
  B} {\bfseries 986} (2023) 116053}.

\bibitem{Langenscheidt:2024nyw}
S.~Langenscheidt and D.~Oriti, \emph{{New edge modes and corner charges for
  first-order symmetries of 4D gravity}},
  \href{https://doi.org/10.1088/1361-6382/adbfee}{\emph{Class. Quant. Grav.}
  {\bfseries 42} (2025) 075010}
  [\href{https://arxiv.org/abs/2408.01809}{{\ttfamily 2408.01809}}].

\bibitem{ciambelli_quantum_2024}
L.~Ciambelli, J.~Kowalski-Glikman and L.~Varrin, \emph{{Quantum corner
  symmetry: Representations and gluing}},
  \href{https://doi.org/10.1016/j.physletb.2025.139544}{\emph{Phys. Lett. B}
  {\bfseries 866} (2025) 139544}
  [\href{https://arxiv.org/abs/2406.07101}{{\ttfamily 2406.07101}}].

\bibitem{freidel_weyl_2021}
L.~Freidel, R.~Oliveri, D.~Pranzetti and S.~Speziale, \emph{The {Weyl} {BMS}
  group and {Einstein}’s equations},
  \href{https://doi.org/10.1007/JHEP07(2021)170}{\emph{Journal of High Energy
  Physics} {\bfseries 2021} (2021) 170}.

\bibitem{varrin_physical_2024}
L.~Varrin, \emph{{Physical representations of corner symmetries}},
  \href{https://doi.org/10.1103/PhysRevD.111.086003}{\emph{Phys. Rev. D}
  {\bfseries 111} (2025) 086003}
  [\href{https://arxiv.org/abs/2409.10624}{{\ttfamily 2409.10624}}].

\bibitem{randers_asymmetrical_1941}
G.~Randers, \emph{On an {Asymmetrical} {Metric} in the {Four}-{Space} of
  {General} {Relativity}},
  \href{https://doi.org/10.1103/PhysRev.59.195}{\emph{Physical Review}
  {\bfseries 59} (1941) 195}.

\bibitem{papapetrou_champs_1966}
A.~Papapetrou, \emph{Champs gravitationnels stationnaires à symétrie axiale},
  {\emph{Annales de l'institut Henri Poincaré. Section A, Physique Théorique}
  {\bfseries 4} (1966) 83}.

\bibitem{bargmann_unitary_1954}
V.~Bargmann, \emph{On {Unitary} {Ray} {Representations} of {Continuous}
  {Groups}}, \href{https://doi.org/10.2307/1969831}{\emph{Annals of
  Mathematics} {\bfseries 59} (1954) 1}.

\bibitem{mackey_unitary_1958}
G.W.~Mackey, \emph{Unitary representations of group extensions. {I}},
  \href{https://doi.org/10.1007/BF02392428}{\emph{Acta Mathematica} {\bfseries
  99} (1958) 265}.

\bibitem{mackey_theory_1976}
G.W.~Mackey, \emph{The {Theory} of {Unitary} {Group} {Representations}},
  University of Chicago Press (1976).

\bibitem{stonevonneumann}
J.~v.~Neumann, \emph{Über einen satz von herrn m. h. stone}, {\emph{Annals of
  Mathematics} {\bfseries 33} (1932) 567}.

\bibitem{kitaev_notes_2018}
A.~Kitaev, \emph{Notes on $\widetilde{SL(2,\mathbb R)}$ representations},
  Aug., 2018.
\newblock 10.48550/arXiv.1711.08169.

\bibitem{lang_discrete_1985}
S.~Lang, \emph{Discrete {Series}},  in \emph{{SL2}({R})}, S.~Lang, ed., (New
  York, NY), pp.~179--190, Springer (1985),
  \href{https://doi.org/10.1007/978-1-4612-5142-2_9}{DOI}.

\bibitem{woodhouse_geometric_1992}
N.M.J.~Woodhouse, \emph{Geometric quantization} (1992).

\bibitem{rubilar_adjoint_2020}
F.~Rubilar and L.~Schultz, \emph{Adjoint orbits of sl(2,r) and their geometry},
   Apr., 2020.
\newblock 10.48550/arXiv.2004.12180.

\bibitem{casselman_representations_nodate}
B.~Casselman, \emph{Representations of {SL2}({R})}, .

\bibitem{AuslanderKonstant}
L.~Auslander and B.~Kostant, ``Polarization and unitary representations of
  solvable lie groups.''

\bibitem{duflo_sur_1977}
M.~Duflo, \emph{Sur la classification des idéaux primitifs dans l'algèbre
  enveloppante d'une algèbre de lie semi-simple},
  \href{https://doi.org/10.2307/1971027}{\emph{Annals of Mathematics}
  {\bfseries 105} (1977) 107}.

\bibitem{knapp}
A.W.~Knapp, ``Representation theory of semisimple groups: An overview based on
  examples.''

\bibitem{lahlali_coset_2024}
I.A.~Lahlali and J.A.~O'Connor, \emph{Coset symmetries and coadjoint orbits},
  Nov., 2024.
\newblock 10.48550/arXiv.2411.05918.

\bibitem{oblak_bms_2017}
B.~Oblak, \emph{{BMS} {Particles} in {Three} {Dimensions}} (2017),
  \href{https://doi.org/10.1007/978-3-319-61878-4}{10.1007/978-3-319-61878-4}.

\bibitem{bates_weinstein_quantization}
S.~Bates and A.~Weinstein, \emph{Lectures on the Geometry of Quantization},
  vol.~8 of \emph{Berkeley Mathematics Lecture Notes}, American Mathematical
  Society and Center for Pure and Applied Mathematics, University of
  California, Berkeley (1997).

\bibitem{Blau}
M.~Blau, ``{Symplectic geometry and geometric quantization}.''
  \url{http://www.blau.itp.unibe.ch/Lecturenotes.html}.

\bibitem{carosso2018geometricquantization}
A.~Carosso, \emph{Geometric quantization},  2018.

\bibitem{lang_sl2r_1985}
S.~Lang, \emph{{SL2}({R})}, vol.~105 of \emph{Graduate {Texts} in
  {Mathematics}}, Springer, New York, NY (1985),
  \href{https://doi.org/10.1007/978-1-4612-5142-2}{10.1007/978-1-4612-5142-2}.

\bibitem{witten_coadjoint_1988}
E.~Witten, \emph{Coadjoint {Orbits} of the {Virasoro} {Group}},
  \href{https://doi.org/10.1007/BF01218287}{\emph{Commun. Math. Phys.}
  {\bfseries 114} (1988) 1}.

\bibitem{plyushchay_quantization_1993}
M.S.~Plyushchay, \emph{Quantization of the classical {SL}(2,{R}) system and
  representations of {SL}(2,{R}) group},
  \href{https://doi.org/10.1063/1.530016}{\emph{J. Math. Phys.} {\bfseries 34}
  (1993) 3954}.

\bibitem{alekseev_quantization_1988}
A.~Alekseev, L.~Faddeev and S.~Shatashvili, \emph{Quantization of symplectic
  orbits of compact {Lie} groups by means of the functional integral},
  \href{https://doi.org/10.1016/0393-0440(88)90031-9}{\emph{Journal of Geometry
  and Physics} {\bfseries 5} (1988) 391}.

\bibitem{Rai:1989js}
B.~Rai and V.G.J.~Rodgers, \emph{{From Coadjoint Orbits to Scale Invariant
  {WZNW} Type Actions and 2-$D$ Quantum Gravity Action}},
  \href{https://doi.org/10.1016/0550-3213(90)90264-E}{\emph{Nucl. Phys. B}
  {\bfseries 341} (1990) 119}.

\bibitem{Brensinger:2017gtb}
S.~Brensinger and V.G.J.~Rodgers, \emph{{Dynamical Projective Curvature in
  Gravitation}}, \href{https://doi.org/10.1142/S0217751X18502238}{\emph{Int. J.
  Mod. Phys. A} {\bfseries 33} (2019) 1850223}
  [\href{https://arxiv.org/abs/1712.05394}{{\ttfamily 1712.05394}}].

\bibitem{Caputa:2018kdj}
P.~Caputa and J.M.~Magan, \emph{{Quantum Computation as Gravity}},
  \href{https://doi.org/10.1103/PhysRevLett.122.231302}{\emph{Phys. Rev. Lett.}
  {\bfseries 122} (2019) 231302}
  [\href{https://arxiv.org/abs/1807.04422}{{\ttfamily 1807.04422}}].

\bibitem{Erdmenger:2020sup}
J.~Erdmenger, M.~Gerbershagen and A.-L.~Weigel, \emph{{Complexity measures from
  geometric actions on Virasoro and Kac-Moody orbits}},
  \href{https://doi.org/10.1007/JHEP11(2020)003}{\emph{JHEP} {\bfseries 11}
  (2020) 003} [\href{https://arxiv.org/abs/2004.03619}{{\ttfamily
  2004.03619}}].

\bibitem{harishchandra_harmonic_1977}
Harish-Chandra, \emph{Harmonic Analysis on Real Reductive Groups}, vol.~576 of
  \emph{Lecture Notes in Mathematics}, Springer-Verlag, Berlin, New York
  (1977).

\bibitem{Ciambelli2025}
L.~Ciambelli, J.~Kowalski-Glikman and L.~Varrin, \emph{Entanglement entropy of
  quantum corners},  2025, \textit{To appear}.

\bibitem{berndt_elements_1998}
R.~Berndt and R.~Schmidt, \emph{Elements of the {Representation} {Theory} of
  the {Jacobi} {Group}}, vol.~163 of \emph{Progress in {Mathematics}},
  Birkhäuser, Basel (1998),
  \href{https://doi.org/10.1007/978-3-0348-8772-4}{10.1007/978-3-0348-8772-4}.

\bibitem{Alekseev:1988ce}
A.~Alekseev and S.L.~Shatashvili, \emph{{Path Integral Quantization of the
  Coadjoint Orbits of the Virasoro Group and 2D Gravity}},
  \href{https://doi.org/10.1016/0550-3213(89)90130-2}{\emph{Nucl. Phys. B}
  {\bfseries 323} (1989) 719}.

\bibitem{Alekseev:1988vx}
A.~Alekseev, L.D.~Faddeev and S.L.~Shatashvili, \emph{{Quantization of
  symplectic orbits of compact Lie groups by means of the functional
  integral}}, \href{https://doi.org/10.1016/0393-0440(88)90031-9}{\emph{J.
  Geom. Phys.} {\bfseries 5} (1988) 391}.

\bibitem{Aratyn:1990dj}
H.~Aratyn, E.~Nissimov, S.~Pacheva and A.H.~Zimerman, \emph{{SYMPLECTIC ACTIONS
  ON COADJOINT ORBITS}},
  \href{https://doi.org/10.1016/0370-2693(90)90420-B}{\emph{Phys. Lett. B}
  {\bfseries 240} (1990) 127}.

\bibitem{Charles:1999qq}
L.~Charles, \emph{{Feynman path integral and Toeplitz quantization}},
  {\emph{Helv. Phys. Acta} {\bfseries 72} (1999) 341}.

\end{thebibliography}\endgroup

\end{document}